\documentclass[a4paper,11pt]{article}
\usepackage[skins]{tcolorbox}
\usepackage{mathtools,slashed}
\usepackage[T1]{fontenc}
\usepackage{slashed,verbatim,subfigure}
\usepackage[numbers,sort&compress]{natbib}
\usepackage{amsmath}
\usepackage{mathrsfs}
\usepackage{amsbsy}
\usepackage{amssymb}
\usepackage{appendix}
\usepackage{graphicx}
\usepackage[final]{pdfpages}
\usepackage{dcolumn}
\usepackage{bm}
\usepackage[colorlinks=true,linktocpage=true,
linkcolor=blue,citecolor=blue]{hyperref}
\usepackage[a4paper]{geometry}
\catcode`@11
\def\seceqaa{\@addtoreset{equation}{section}
	\def\theequation{A\arabic{equation}}}
\def\seceqbb{\@addtoreset{equation}{section}
	\def\theequation{B\arabic{equation}}}
\def\seceqcc{\@addtoreset{equation}{section}
	\def\theequation{C\arabic{equation}}}
\def\seceqdd{\@addtoreset{equation}{section}
	\def\theequation{D\arabic{equation}}}
\def\seceqee{\@addtoreset{equation}{section}
	\def\theequation{E\arabic{equation}}}
\catcode`@11
\newcommand{\be}{\begin{eqnarray}}
\newcommand{\ee}{\end{eqnarray}}
\begin{document}
\large
\title{Multiverse in Karch-Randall Braneworld}
\author{\hspace{-0.5cm}Gopal Yadav\footnote{email- gyadav@ph.iitr.ac.in}\vspace{0.1in}\\
Department of Physics,\\
Indian Institute of Technology Roorkee, Roorkee 247667, Uttarakhand, India}
\date{}
\maketitle

\begin{abstract}
In this paper, we propose a model based on wedge holography that can describe the multiverse. In wedge holography, we consider two gravitating baths, one of which has strong gravity and the other one has weak gravity. To describe a multiverse, we consider $2n$ Karch-Randall branes, and we propose that various $d$-dimensional universes are localized on these branes. These branes are embedded in $(d+1)$-dimensional spacetime. The model is useful in obtaining the Page curve of black holes with multiple horizons and in the resolution of the ``grandfather paradox''. We explicitly obtain the Page curves of eternal AdS black holes for $n=2$ multiverse and Schwarzschild de-Sitter black hole with two horizons.
\end{abstract}

\newpage	
\tableofcontents
\section{Introduction}
Recently doubly holographic setup has drawn the attention of many researchers to study the information paradox \cite{Hawking}. A version of the resolution of information paradox is to get the Page curve \cite{Page}. AdS/CFT conjecture states that bulk gravity is dual to quantum field theory on the AdS boundary \cite{AdS-CFT}. Doubly holographic setup is the extended version where one considers two copies of AdS/BCFT-like systems \cite{PBD,Bath-WCFT,B-N-P-1,B-N-P-2,NGB,Ling+Liu+Xian,Omiya+Wei,RE-DH,Phase-BCFT,critical-islands,IITK1,IITK2,Liu et al,Li+Yang,Deng+An+Zhou,Island-IIB-1,Island-IIB-2,Island-IIB-3,Geng+Nomura+Sun,Geng-2,HD-Page Curve-2,JIITK}. The idea was started from the Karch-Randall model, where one chop off the AdS boundary by a Karch-Randall brane \cite{KR1,KR2}. Let us discuss three equivalent descriptions of the doubly holographic setup which is being used to obtain the Page curve.
\begin{itemize}
\item BCFT is living on $d$-dimensional boundary of AdS spacetime. BCFT has a $(d-1)$-dimensional boundary, known as a defect.

\item Gravity on $d$-dimensional Karch-Randall brane is coupled to BCFT at the defect via transparent boundary condition.

\item $d$-dimensional BCFT has gravity dual which is Einstein gravity on $AdS_{d+1}$.
\end{itemize}
In this setup, the Karch-Randall brane contains a black hole whose Hawking radiation is collected by BCFT bath. One can define the radiation region on the BCFT bath, and the entanglement entropy of Hawking radiation can be obtained using the semiclassical formula in the second description \cite{AMMZ}. The advantage of a doubly holographic setup is that we can compute entanglement entropy very easily using the classical Ryu-Takayanagi formula \cite{RT} in the third description. In this kind of setup, there exist two types of extremal surfaces: Hartman-Maldacena surface \cite{Hartman-Maldacena}, which starts at the defect, crosses the black hole horizon, and goes to its thermofield double partner; in this process volume of Einstein-Rosen bridge grows. Another extremal surface is the island surface, which starts at BCFT and lands on the Karch-Randall brane. It turns out that initially, the entanglement entropy of the Hartman-Maldacena surface dominates, and after the Page time island surface takes over, and hence one gets the Page curve. The problem with this setup is that gravity becomes massive on the Karch-Randall brane, which is not physical \cite{massive-gravity,GB-2,GB-3,GB-4}. See \cite{Massless-Gravity,critical-islands,HD-Page Curve-2,C1} for computation of Page curve with massless gravity on Karch-Randall brane. Massless gravity on Karch-Randall brane in \cite{Massless-Gravity} arises due to the inclusion of the Dvali-Gabadadze-Porrati term \cite{DGP-2} on the same. In \cite{HD-Page Curve-2}, we explicitly showed that normalizable graviton wave function requires massless graviton. Another reason is that we implemented the Dirichlet boundary condition on the graviton wave function at the black hole horizon that quantized the graviton mass and allowed a massless graviton. Further, the tension of the Karch-Randall brane (in our case it was a fluxed hyper-surface) is inversely proportional to the black hole horizon and we obtained ``volcano''-like potential hence one can localize massless gravity on the Karch-Randall brane. Despite massless gravity on the Karch-Randall brane, we had comparable entanglement entropies coming from Hartman-Maldacena and island surfaces. Therefore we obtained the Page curve of an eternal neutral black hole from a top-down approach. In \cite{C1}, authors imposed Dirichlet boundary conditions on gravitating branes in wedge holography where they obtained the Page curve even in the presence of massless gravity. The existence of islands with massless gravity was present in \cite{critical-islands} because of the geometrical construction of the critical Randall-Sundrum II model. Information paradox of flat space black holes was discussed in \cite{Flat-space-black-holes,Flat-space-black-holes-2,Flat-space-black-holes-3}\footnote{We thank C.~Krishnan for pointing out these interesting papers to us.} where one defines the subregions on the holographic screen to compute holographic entanglement entropy. The setup in which the bath is also gravitating is known as ``wedge holography'' \cite{WH-1,WH-2,WH-iii}. See \cite{S1,S2,S3,S4} for work on quantum entanglement, complexity, and entanglement negativity in de-Sitter space\footnote{We thank S.~Choudhury to bring his works to our attention.}.

In wedge holography, we consider two Karch-Randall branes, $Q_1$ and $Q_2$, of tensions $T_1$ and $T_2$ such that $T_1<T_2$. In this setup, $Q_2$ contains a black hole whose Hawking radiation is collected by $Q_1$. Literature on wedge holography can be found in \cite{WH-i,WH-ii,WH-iii,Geng}.

It is easy to obtain the Page curve for black holes with a single horizon. In this paper, we address the following issues: we construct a multiverse using the idea of wedge holography and use this setup to get the Page curve of black holes with multiple horizons from wedge holography. Multiverse in this paper will be constructed by localizing Einstein's gravity on various Karch-Randall branes. These branes will be embedded into one higher dimension. Further, we propose that it is possible to travel between different universes because all are communicating with each other. We suspect that the ``grandfather paradox'' can be resolved in this setup.

The paper is organized as follows. In section \ref{review-WH}, we briefly review wedge holography. In section \ref{Multiverse-section}, we discuss the existence of multiverse in the Karch-Randall braneworld with geometry anti de-Sitter, de-Sitter, and the issues when we mix de-Sitter and anti de-Sitter spacetimes in subsections \ref{AdS-multiverse}, \ref{de-Sitter-multiverse} and \ref{AdS+dS-Multiverse}. In sections \ref{AIP}, we discuss application of the multiverse to information paradox where we have obtained the Page curve of eternal AdS black holes for $n=2$ multiverse in \ref{PC-TEBH} and Page curve of Schwarzschild de-Sitter black hole in \ref{IP-SdS} via \ref{Sch-patch} and \ref{dS-patch}. Section \ref{AGP} is on the application of this model to grandfather paradox. Finally, we discuss our results in section \ref{Conclusion}.

\section{Brief Review of Wedge Holography}
\label{review-WH}
In this section, let us review wedge holography \cite{WH-1,WH-2,WH-iii}. Consider the following action.
\begin{eqnarray}
\label{bulk-action}
S=-\frac{1}{16 \pi G_N^{(d+1)}} \int d^{d+1}x \sqrt{-g_{\rm bulk}} \left(R[g_{\rm bulk}] - 2 \Lambda\right)-\frac{1}{8 \pi G_N^{(d+1)}} \int_{\alpha=1,2} d^dx \sqrt{-h_\alpha}\left({\cal K}_\alpha-T_\alpha\right),
\end{eqnarray}
where first term is the Einstein-Hilbert term with negative cosmological constant$\left(\Lambda=-\frac{d(d-1)}{2}\right)$, and second term correspond to boundary terms on Karch-Randall branes of tensions $T_{\alpha=1,2}$.
Einstein equation for the bulk action (\ref{bulk-action}) turns out to be:
\begin{equation}
\label{Einstein-equation}
R_{\mu \nu}-\frac{1}{2}g_{\mu \nu} R =\frac{d(d-1)}{2} g_{\mu \nu}.
\end{equation}
Solution to Einstein equation is \cite{WH-2}:
\begin{equation}
\label{metric-bulk}
ds_{(d+1)}^2=g_{\mu \nu} dx^\mu dx^\nu=dr^2+\cosh^2(r) h_{ij}^{\alpha} dy^i dy^j,
\end{equation}
where $h_{ij}^{\alpha}$ are the induced metric on Karch-Randall branes.
\begin{figure}
\begin{center}
\includegraphics[width=0.9\textwidth]{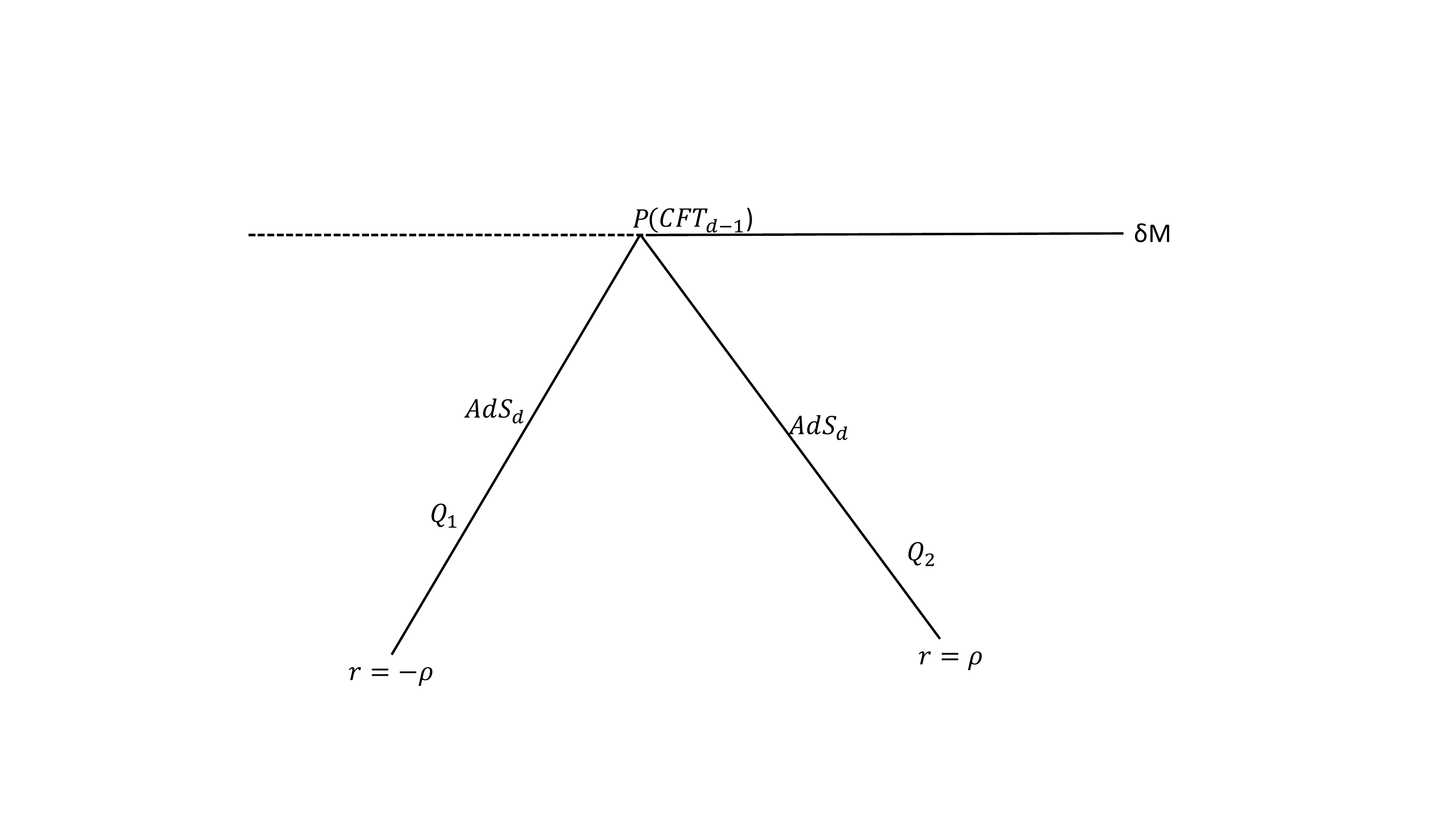}
\end{center}
\caption{Description of wedge holography. Two $d$-dimensional Karch-Randall branes joined at the $(d-1)$ dimensional defect, Karch-Randall branes are embedded in $(d+1)$-dimensional bulk. }
\label{WHS}
\end{figure}
One can obtain Neumann boundary condition by the variation of (\ref{bulk-action}) with respect to $h_{ij}^{\alpha}$ and is given as: 
\begin{equation}
\label{NBC}
{\cal K}_{ij}^{\alpha}-({\cal K}^{\alpha}-T^{\alpha})h_{ij}^{\alpha}=0.
\end{equation}
For the consistent construction of wedge holography, the metric (\ref{metric-bulk}) should be the solution of (\ref{Einstein-equation}) provided $h_{ij}^{\alpha}$ should satisfy Einstein equation with a negative cosmological constant in d dimensions
\begin{eqnarray}
\label{Brane-Einstein-equation}
R_{ij}^{\alpha}-\frac{1}{2}h_{ij}^{\alpha} R[h_{ij}]^{\alpha} =\frac{(d-1)(d-2)}{2} h_{ij}^{\alpha},
\end{eqnarray}
and it should satisfy Neumann boundary condition (\ref{NBC}) at $r=\pm \rho$. See Fig. \ref{WHS} for a pictorial representation of wedge holography.
 One can also choose $-\rho_1 \leq r \leq \rho_2$ with $\rho_1 \neq \rho_2$ \cite{WH-2}, in this range, tensions of the branes will be different. This is useful in obtaining the Page curve. There are three descriptions of wedge holography summarised below:
\begin{itemize}
\item {\bf Boundary description:} $CFT_{d-1}$ living on the wedge of common boundaries of two $AdS_d$'s.

\item {\bf Intermediate description:}
Two Karch-Randall branes of geometry $AdS_d$ ($Q_1$ and $Q_2$) glued to each other at the interface point by a transparent boundary condition.

\item {\bf Bulk description:} Einstein gravity in $(d+1)$-dimensional bulk, $AdS_{d+1}$.
\end{itemize}
Precisely, correspondence can be interpreted as:
``Classical gravity in $(d+1)$-dimensions has a holographic dual theory on the defect which is CFT in $(d-1)$-dimensions''. 

Wedge holography is useful in the computation of the Page curve of black holes. Let us understand this connection. In the intermediate description, we consider a black hole on $Q_2$ whose Hawking radiation will be collected by weakly gravitating bath $Q_1$ (i.e., $T_1<T_2$). To calculate the entanglement entropy in the intermediate description, one is required to use the semiclassical formula:
\begin{equation}
S({\cal R}) = {\rm min}_{\cal I}~{\rm ext}_{\cal I}~S_{gen}({\cal R} \cup {\cal I}),
\end{equation}
where \cite{EW}:
\begin{equation}
S_{gen}({\cal R} \cup {\cal I})= \frac{A(\partial {\cal I})}{4 G_N}+ S_{matter}({\cal R} \cup {\cal I}),
\end{equation}
where $A(\partial {\cal I})$ denotes the area of the boundary of the island surface, and $S_{matter}({\cal R} \cup {\cal I})$ interpreted as matter contributions from radiation and island regions both. Using bulk description, we can obtain entanglement entropy using the classical Ryu-Takayanagi formula \cite{RT}.
\begin{equation}
S_{gen}({\cal R} \cup {\cal I})= \frac{A(\gamma)}{4 G_N^{(d+1)}},
\end{equation}
where $\gamma$ is the minimal surface in bulk. In wedge holography, there is one more extremal surface, Hartman-Maldacena surface \cite{Hartman-Maldacena}, which starts at the defect, crosses the horizons, and meets its thermofield double. By plotting the entanglement entropies contributions of these surfaces, we can get the Page curve \cite{Page}.

\section{Emerging Multiverse from Wedge Holography}
\label{Multiverse-section}
In this section, we discuss how one can describe multiverse from wedge holography.

\subsection{Anti de-Sitter Background}
\label{AdS-multiverse}
In this subsection, we construct a multiverse from $AdS$ spacetimes. Let us first start with the simplest case discussed in \ref{review-WH}. To describe multiverse, we need multiple Karch-Randall branes located at $r=\pm n \rho$ such that bulk metric should satisfy Neumann boundary condition at the aforementioned locations. Extrinsic curvature on the Karch-Randall brane and its trace is computed as:
\begin{eqnarray}
\label{Extrinsic-Curvature}
& & 
{\cal K}_{ij}^\alpha=  \frac{1}{2} \left(\partial_r g_{ij}\right)|_{r=\pm n \rho} = \tanh( r) g_{ij}|_{r=\pm n \rho} =\tanh(\pm n \rho) h_{ij}^\alpha ,\nonumber\\
& & {\cal K}^\alpha=h^{ij}_\alpha K_{ij}^\alpha= d \tanh(\pm n \rho).
\end{eqnarray}
We can see that Neumann boundary condition (\ref{NBC}) is satisfied at $r=\pm n \rho$ provided $T_{\rm AdS}^\alpha=(d-1) \tanh(\pm n \rho)$\footnote{It seems that some of branes have negative tension. Let us discuss the case when branes are located at $- n \rho_1$ and $ n \rho_2$ with $\rho_1 \neq \rho_2$. In this case tensions of branes are $(d-1) \tanh(- n \rho_1)$ and $(d-1) \tanh( n \rho_2)$. Negative tension issue can be resolved when we consider $\rho_1 <0$ and $\rho_2>0$ similar to \cite{WH-iii}. Therefore this fixes the brains stability issue in our setup. This discussion is also applicable to the case when $\rho_1=\rho_2$.}, where $\alpha=-n,...,n$. Further, bulk metric (\ref{metric-bulk}) is also satisfying the Einstein equation (\ref{Einstein-equation}), and hence, this guarantees the existence of $2n$ Karch-Randall branes in our setup. These $2n$-branes are analogs of universes that are embedded in $AdS_{d+1}$. Defect is described as: $P=Q_\alpha \cap Q_\beta$, where $\alpha,\beta=-n, -n+1,..,1,...,n-1,n$. Now, we include the DGP term in the gravitational action, which can describe massless gravity \cite{Massless-Gravity}.
{\footnotesize
\begin{eqnarray}
\label{bulk-action-DGP}
& &
\hskip -0.2in S=\frac{1}{16 \pi G_N^{(d+1)}}\Biggl[ \int_M d^{d+1}x \sqrt{-g} \left(R[g] + {d(d-1)}\right)+2\int_{\partial M} d^d x\sqrt{-h}K+2 \int_{Q_\alpha} d^dx \sqrt{-h_\alpha}\left({\cal K}_\alpha-T_\alpha+\lambda_\alpha R_{h_\alpha}\right)\Biggr] , 
\end{eqnarray}
}
where the first term is the bulk Einstein-Hilbert term with negative cosmological constant, the second term is the Gibbons-Hawking-York boundary term for conformal boundary $\partial M$, and the third term corresponds to the difference of extrinsic curvature scalar and tensions of $2n$ Karch-Randall branes, $R_{h_\alpha}$ are intrinsic curvature scalars for $2n$ Karch-Randall branes. DGP is understood as the Dvali-Gabadadze-Porrati term \cite{DGP-2}. In this case, bulk metric satisfies the following Neumann boundary condition at $r=\pm n \rho$\footnote{When we discuss multiverse then $\alpha$ and $\beta$ will take $2n$ values whereas when we discuss wedge holography then $\alpha,\beta=1,2$. }
\begin{eqnarray}
\label{NBC-DGP}
{\cal K}_{\alpha,ij}-({\cal K}_\alpha-T_\alpha+\lambda_\alpha R_{h_\alpha} )h_{\alpha,ij}+2 \lambda_\alpha R_{\alpha, {ij}} =0.
\end{eqnarray}
Einstein equation for the bulk action (\ref{bulk-action-DGP}) will be same as (\ref{Einstein-equation}) and hence solution is:
\begin{equation}
\label{metric-bulk-DGP}
ds_{(d+1)}^2=g_{\mu \nu} dx^\mu dx^\nu=dr^2+\cosh^2(r) h_{ij}^{\alpha, \rm AdS} dy^i dy^j,
\end{equation}
with $-n \rho_1 \leq r \leq n \rho_2$. Induced metric $h_{ij}^\alpha$ satisfy Einstein equation on the brane
\begin{eqnarray}
\label{Brane-Einstein-equation-DGP}
R_{ij}^\alpha-\frac{1}{2}h_{ij}^\alpha R[h_{ij}]^\alpha =\frac{(d-1)(d-2)}{2} h_{ij}^\alpha.
\end{eqnarray}
The above equation can be derived from the following Einstein-Hilbert term including negative cosmological constant on the brane:
\begin{eqnarray}
\label{boundary-EH-action-AdS}
S_{\rm AdS}^{\rm EH} =\lambda_\alpha^{\rm AdS} \int d^{d}x \sqrt{-h_{\alpha}} \left(R[h_{\alpha}] - 2 \Lambda_{\rm brane}^{\rm AdS}\right),
\end{eqnarray}
where $\Lambda_{\rm brane}^{\rm AdS}=-\frac{(d-1)(d-2)}{2}$, $\lambda_\alpha^{\rm AdS}  \left(\equiv \frac{1}{16 \pi G_N^{d,\ \alpha}}=\frac{1}{16 \pi G_N^{(d+1)}}\int_0^{\alpha \rho}\cosh^{d-2}(r) dr \ ; (\alpha=1,2,...,n)\right)$\footnote{Explicit derivation of (\ref{boundary-EH-action-AdS}) was done in \cite{WH-2} for two Karch-Randall branes. One can generalize the same for $2n$ Karch-Randall branes. In this setup upper limit of integration will be different for different locations of Karch-Randall branes.} is related to effective Newton's constant in $d$ dimensions, and (\ref{boundary-EH-action-AdS}) can be obtained by substituting (\ref{metric-bulk-DGP}) into (\ref{bulk-action}) and using the value of ${\cal K}^{\alpha}$ from (\ref{Extrinsic-Curvature}) and branes tensions  $T_{\rm AdS}^\alpha=(d-1) \tanh(\pm n \rho)$.
\begin{figure}
\begin{center}
\includegraphics[width=0.8\textwidth]{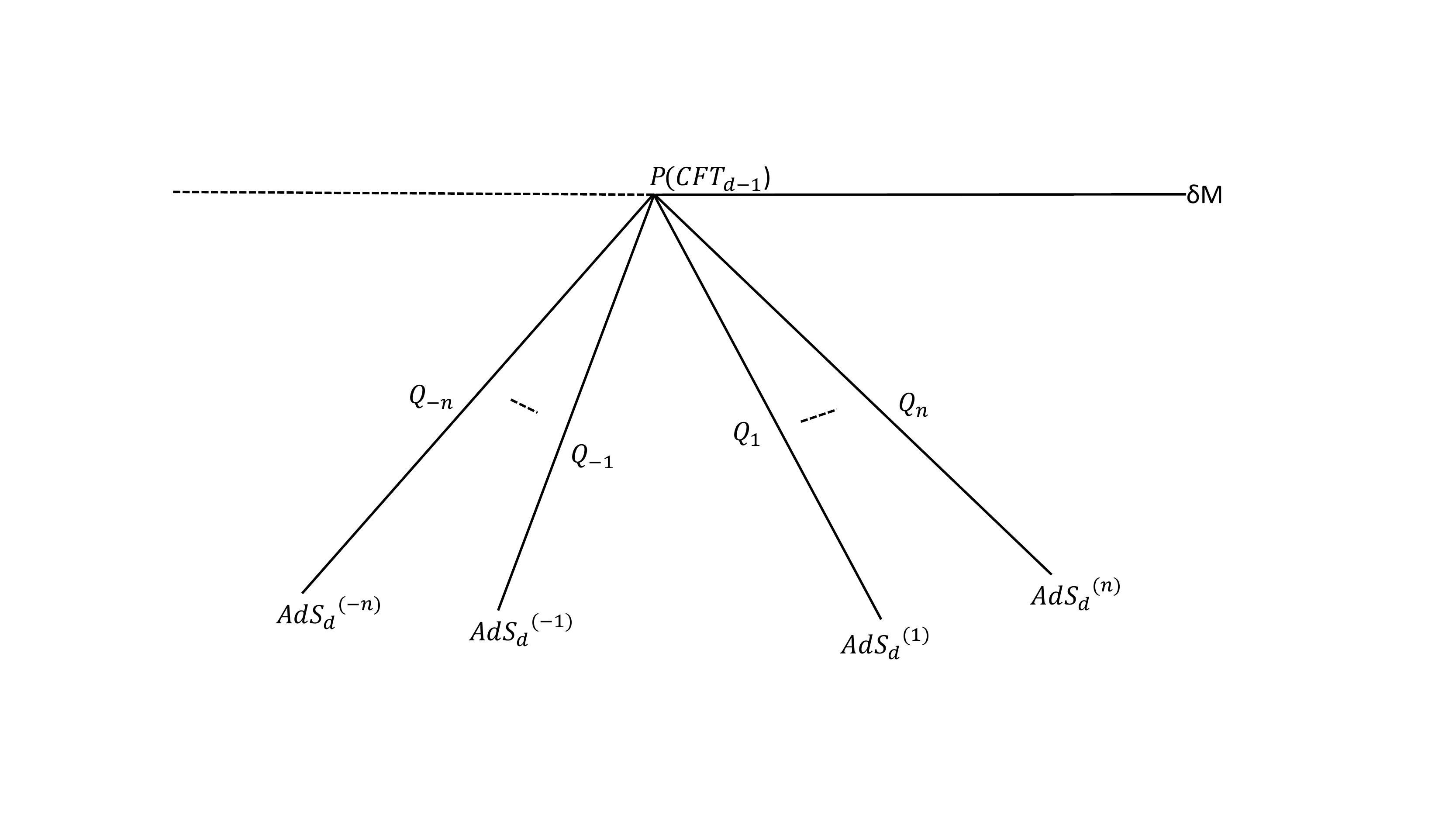}
\end{center}
\caption{$2n$ Karch-Randall branes, $Q_{-n,-n+1,...,1,2,...,n-1,n}$ embedded in $AdS_{d+1}$. P is the defect. Multiverse is described by $2n$ Karch-Randall branes which are $d$-dimensional objects and defect is $(d-1)$-dimensional object.}
\label{Multiverse-AdS-i}
\end{figure}

Three descriptions of our setup are as follows:
\begin{itemize}
\item {\bf Boundary description:} $d$-dimensional boundary conformal field theory with $(d-1)$-dimensional boundary.

\item {\bf Intermediate description:} All $2n$ gravitating systems are connected at the interface point by transparent boundary condition.

\item {\bf Bulk description:} Einstein gravity in the $(d+1)$-dimensional bulk.

\end{itemize}
\begin{figure}
\begin{center}
\includegraphics[width=0.8\textwidth]{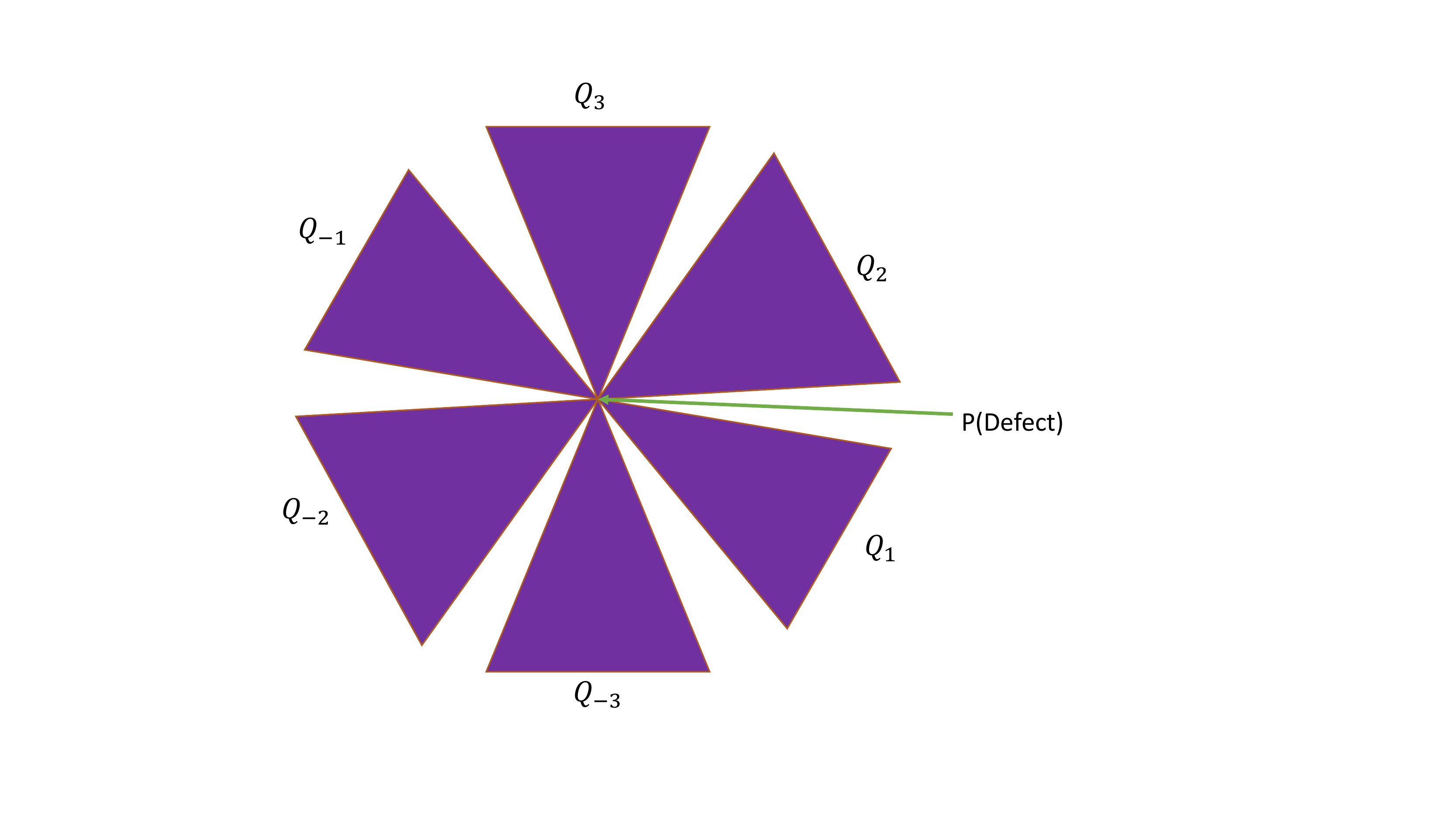}
\end{center}
\caption{Cartoon picture of the multiverse for $n=3$ in AdS spacetimes. P is the $(d-1)$-dimensional defect and Karch-Randall branes are denoted by $Q_{-1/1,-2/2,-3/3}$.}
\label{CP-AdS-Multiverse}
\end{figure}
We see that in the intermediate description, there is a transparent boundary condition at the defect; therefore multiverse constructed in this setup consists of communicative universes localized on Karch-Randall branes (see Figs. \ref{Multiverse-AdS-i},\ref{CP-AdS-Multiverse}). Wedge holography dictionary for ``multiverse'' with $2n$ AdS branes can be stated as follows.\\ \\
\fbox{\begin{minipage}{38em}{\it 
Classical gravity in $(d+1)$-dimensional anti de-Sitter spacetime\\ $\equiv$ (Quantum) gravity on $2n$ $d$-dimensional Karch-Randall branes with metric $AdS_d$\\ $\equiv$ CFT living on $(d-1)$-dimensional defect.}
\end{minipage}}\\ \\
 Second and third line exist due to braneworld holography \cite{KR1,KR2} and usual AdS/CFT correspondence \cite{AdS-CFT} due to gravity on the brane. Therefore, {\it classical gravity in $AdS_{d+1}$ is dual to $CFT_{d-1}$ at the defect.}

\subsection{de-Sitter Background}
\label{de-Sitter-multiverse}
In this subsection, we study the realization of the multiverse in such a way that the geometry of Karch-Randall branes is of de-Sitter spacetime. Wedge holography with de-Sitter metric on Karch-Randall branes was discussed in \cite{WH-2} where the bulk spacetime is AdS spacetime and in \cite{FS-Holography} with flat space bulk metric. Before going into the details of construction of ``multiverse'' with de-Sitter geometry on Karch-Randall branes, first let us summarise some key points of \cite{FS-Holography}.

Authors in \cite{FS-Holography} constructed wedge holography in $(d+1)$-dimensional flat spacetime with Lorentzian signature. Karch-Randall branes in their construction have either geometry of $d$-dimensional hyperbolic space or de-Sitter space. Since our interest lies in the de-Sitter space therefore we only discuss the results related to the same. Geometry of the defect is $S^{d-1}$. Wedge holography states that

\fbox{\begin{minipage}{38em}{\it 
Classical gravity in $(d+1)$-dimensional flat spacetime\\ $\equiv$ (Quantum) gravity on two $d$-dimensional Karch-Randall branes with metric $dS_d$\\ $\equiv$ CFT living on $(d-1)$-dimensional defect $S^{d-1}$.}
\end{minipage}}\\ \\
 Third line in the above duality is coming from dS/CFT correspondence \cite{dS-CFT, dS-CFT-1}. Authors in \cite{FS-Holography} explicitly calculated the central charge of dual CFT which was imaginary and hence CFT living at the defect is non-unitary. \par
 The above discussion also applies to the AdS bulk as well. In this case one can state the wedge holographic dictionary as:\\ \\
\fbox{\begin{minipage}{38em}{\it 
Classical gravity in $(d+1)$-dimensional anti de-Sitter spacetime\\ $\equiv$ (Quantum) gravity on two $d$-dimensional Karch-Randall branes with metric $dS_d$\\ $\equiv$ non-unitary CFT living at the $(d-1)$-dimensional defect.}
\end{minipage}}\\

\begin{figure}
\begin{center}
\includegraphics[width=0.8\textwidth]{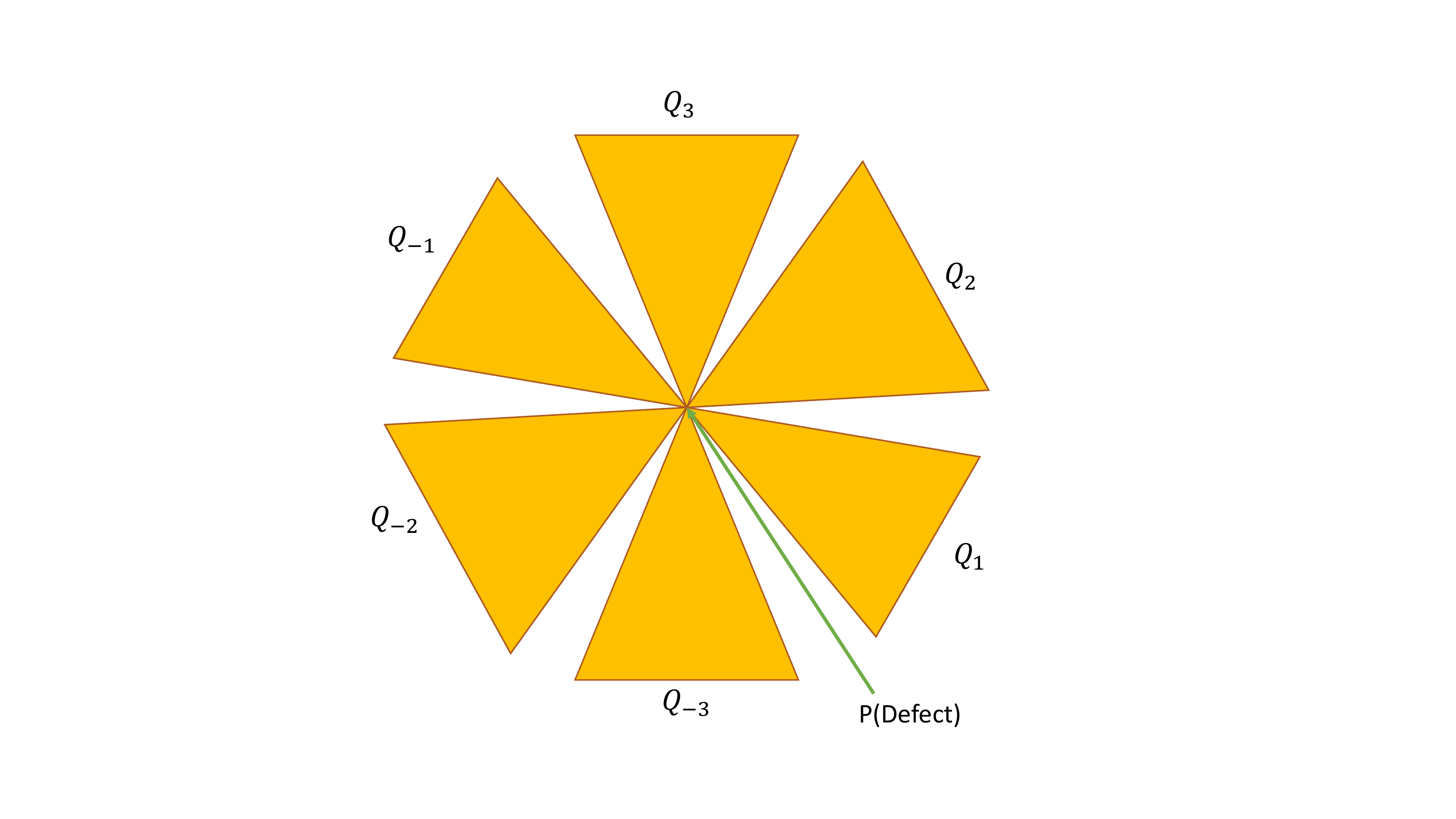}
\end{center}
\caption{Cartoon picture of the multiverse for $n=3$ with de-Sitter metric on Karch-Randall branes. P is the $(d-1)$-dimensional defect and Karch-Randall branes are denoted by $Q_{-1/1,-2/2,-3/3}$.}
\label{CP-dS-Multiverse}
\end{figure}

 Now to discuss the existence of multiverse, we start with the bulk metric \cite{WH-2}:
\begin{eqnarray}
\label{de-Sitter-metric}
ds_{(d+1)}^2=g_{\mu \nu} dx^\mu dx^\nu=dr^2+\sinh^2(r) h_{ij}^{\beta, \rm dS} dy^i dy^j,
\end{eqnarray}
(\ref{de-Sitter-metric}) is the solution of (\ref{Einstein-equation}) with a negative cosmological constant provided induced metric on Karch-Randall brane ($h_{ij}^\beta$) is the solution of Einstein equation with a positive cosmological constant on Karch-Randall branes:
\begin{eqnarray}
\label{Brane-Einstein-equation-dS}
R_{ij}^\beta-\frac{1}{2}h_{ij}^\beta R[h_{ij}]^\beta =-\frac{(d-1)(d-2)}{2} h_{ij}^\beta.
\end{eqnarray}
One can derive Einstein-Hilbert terms with positive cosmological constant on Karch-Randall branes by using Neumann boundary condition (\ref{NBC}) for de-Sitter branes and substituting (\ref{de-Sitter-metric}) in (\ref{bulk-action}), the resulting action is given by the following expression
\begin{eqnarray}
\label{boundary-EH-action-dS}
S_{\rm dS}^{\rm EH} =\lambda_\beta^{\rm dS} \int d^{d}x \sqrt{-h_{\beta}} \left(R[h_{\beta}] - 2 \Lambda_{\rm brane}^{\rm dS}\right),
\end{eqnarray}
where  $\lambda_\beta^{\rm dS}\left(\equiv \frac{1}{16 \pi G_N^{d,\ \beta}}=\frac{1}{16 \pi G_N^{(d+1)}}\int_0^{\beta \rho}\sinh^{d-2}(r) dr \ ; (\beta=1,2,...,n)\right)$\footnote{See \cite{WH-2} for the explicit derivation. Only difference is that, in our setup, we have $\beta=1,2,...,n$.} represents relationship with effective Newton's constant on the branes and $\Lambda_{\rm brane}^{\rm dS}=\frac{(d-1)(d-2)}{2}$.
For the de-Sitter embeddings in bulk AdS spacetime (\ref{de-Sitter-metric}), extrinsic curvature and trace of the same on the Karch-Randall branes are obtained as:
\begin{eqnarray}
\label{Extrinsic-Curvature-dS}
& & 
{\cal K}_{ij}^\beta=  \frac{1}{2} \left(\partial_r g_{ij}\right)|_{r=\pm n \rho} = \coth( r) g_{ij}|_{r=\pm n \rho} =\coth(\pm n \rho) h_{ij}^\beta ,\nonumber\\
& & {\cal K}^\beta=h^{ij}_\beta K_{ij}^\beta= d \coth(\pm n \rho).
\end{eqnarray}
Using (\ref{Extrinsic-Curvature-dS}), we can see that (\ref{de-Sitter-metric}) satisfy Neumann boundary condition (\ref{NBC}) at $r=\pm n \rho$ if the tensions of branes are $T_{\rm dS}^{\beta}=(d-1)\coth\left(\pm n\rho\right)$, where $\beta=-n,...,n$. Therefore we can obtain $2 n$ copies of Karch-Randall branes with metric de-Sitter spacetime on each of the brane. In this case, {\it the multiverse consists of $2n$ universes localized on Karch-Randall branes whose geometry is $dS_d$, and these $2n$ copies are embedded in $AdS_{d+1}$}. Pictorial representation of the same for $n=3$ is given in the Fig. \ref{CP-dS-Multiverse}. 
Now let us discuss the three descriptions of multiverse with de-Sitter geometries on Karch-Randall branes.
\begin{itemize}
\item {\bf Boundary description:} $d$-dimensional BCFT with $(d-1)$-dimensional defect.

\item {\bf Intermediate description:} $2n$ gravitating systems with de-Sitter geometry connected to each other at the $(d-1)$-dimensional defect.

\item {\bf Bulk description:} $(d+1)$-dimensional Einstein gravity with negative cosmological constant in the bulk.
\end{itemize}
First and third description are related to each other via AdS/BCFT correspondence and $(d-1)$-dimensional defect which is non-unitary CFT exists because of dS/CFT correspondence \cite{dS-CFT,dS-CFT-1}. de-Sitter space exists for finite time and then disappear. Another de-Sitter space born after the disappearance of previous one \cite{mismatched-branes}. Therefore it is possible to have a ``multiverse'' (say $M_1$) with de-Sitter branes provided all of them should be created at the same ``creation time''\footnote{Creation time is defined as the ``time'' when any universe born \cite{mismatched-branes}.} but this will exist for finite time and then $M_1$ disappears. After disappearance of $M_1$, other multiverse (say $M_2$) consists of many de-Sitter branes  born with same creation of time of all the de-Sitter branes.

\subsection{Braneworld Consists of Anti de-Sitter and de-Sitter Spacetimes}
\label{AdS+dS-Multiverse}
Based on the discussion in \ref{AdS-multiverse} and \ref{de-Sitter-multiverse}, we can construct two copies of multiverse, ${M}_1$ and ${M}_2$ in such a way that metric of Karch-Randall branes in ${M}_1$ have the structure of $AdS_d$ spacetime and Karch-Randall branes in ${M}_2$ have geometry of de-Sitter spaces in $d$-dimensions. Bulk metric (\ref{metric-bulk}) of ${M}_1$ and (\ref{de-Sitter-metric}) of ${M}_2$ satisfy Einstein's equation with a negative cosmological constant in bulk (\ref{Einstein-equation}). In this scenario, ${M}_1$ consists of $2n_1$ Karch-Randall branes located at $r=\pm n_1 \rho$ with induced metric $h_{ij}^{\alpha,\rm AdS}$, and tensions $T_{\rm AdS}^\alpha=(d-1)\tanh(\pm n_1 \rho)$  and ${ M}_2$ contains $2n_2$ Karch-Randall branes located at $r=\pm n_2 \rho$ with induced metric $h_{ij}^{\beta,\rm dS}$, and tensions $T_{\rm dS}^\beta=(d-1)\coth(\pm n_2 \rho)$, where $\alpha=-n_1,...,n_1$ and $\beta=-n_2,...,n_2$.

  One can ask why we are interested in the setup that contains mixture of anti de-Sitter and de-Sitter branes. The answer is that this model will be helpful in the study of information paradox of the Schwarzschild de-Sitter black hole with two horizons from wedge holography. To do so, one has to replace AdS branes in ${M}_1$ with flat-space branes\footnote{In this case, warp factor will be different in the bulk metric. Exact metric is given in (\ref{metric-bulk-flat-space}).} with $n_1=1$. Overall, we have  $n_1=n_2=1$ such that we have two flat-space branes and two de-Sitter branes.

\begin{figure}
\begin{center}
\includegraphics[width=0.8\textwidth]{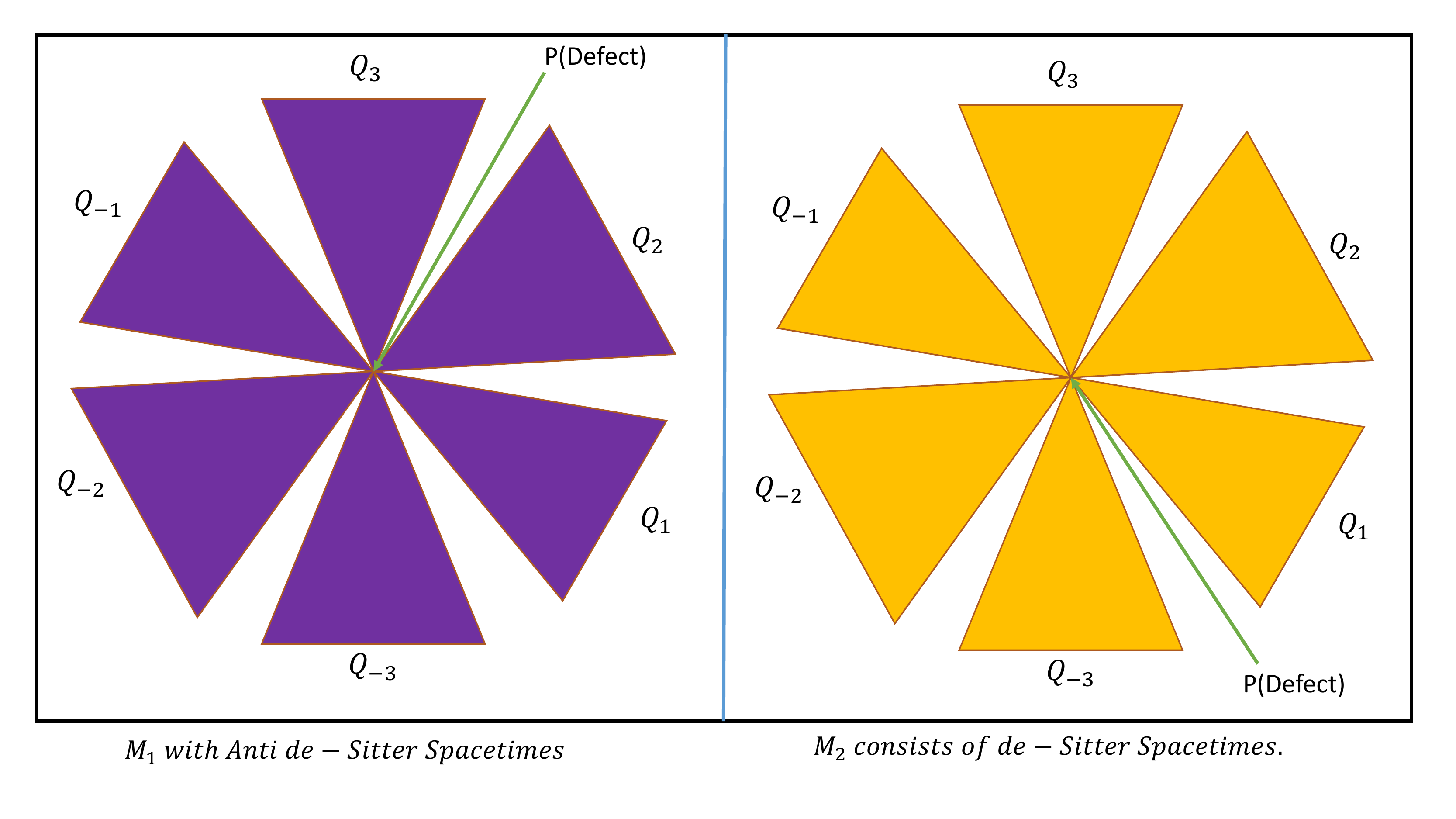}
\end{center}
\caption{Braneworld consists of $d$-dimensional anti de-Sitter and de-Sitter spacetimes. AdS spacetimes are embedded in the bulk (\ref{metric-bulk}) where as de-Sitter spacetimes are embedded in the bulk spacetime with metric (\ref{de-Sitter-metric}). We have used $n_1=n_2=3$ to draw this figure.}
\label{CP-dS+AdS-Multiverse}
\end{figure}

Now the question is that whether this description makes sense or not. When $d$-dimensional AdS spacetimes are embedded in $AdS_{d+1}$ then these branes intersect at the time-like surface of  $AdS_{d+1}$ boundary whereas when $dS_d$ Karch-Randall branes are embedded in $AdS_{d+1}$ then they intersect at the space-like surface of $AdS_{d+1}$ boundary\footnote{We thank J.~Maldacena for comment on this.}. In Fig. \ref{CP-dS+AdS-Multiverse}, as long as $M_1$ and $M_2$ are disconnected from each other then there is no problem. This is what has been followed in \ref{IP-SdS} to get the Page curve of Schwarzschild de-Sitter black hole by treating Schwarzschild and de-Sitter patches independent of each other.

In this subsection, we have discussed the embedding of different types of Karch-Randall branes in the different bulks which are disconnected from each other. Authors in \cite{mismatched-branes} have discussed the various possibilities of embedding of different types of branes, e.g., Minkowski, de-Sitter and anti de-Sitter branes in the same bulk. Existence of various branes are characterized by creation time $\tau_*$. There is finite amount of time for which Minkowski and de-Sitter branes born and there is no cration time for anti de-Sitter branes. Out of various possibilities discussed in \cite{mismatched-branes}, it was pointed by authors that one can see Minkowski, de-Sitter and anti de-Sitter brane at the same time with creation time $\tau_*=-\pi/2$ in a specific bulk. In this case, branes have time dependent position. First we will summarise this result\footnote{For more details, see \cite{mismatched-branes}.} and then comment on the realization of the same from wedge holography.

Bulk $AdS_5$ metric has the following form:
\begin{eqnarray}
\label{metric-AdS5-KR}
ds^2=\frac{1}{z^2}\left(-dt_h^2+t_h^2 dH_3^2+dz^2\right),
\end{eqnarray}
where $dH_3^2=d\theta^2+\sinh^2(\theta)d\omega_2^2$. In this bulk, Minkowski Randall-Sundrum brane is located at $z_M(t_h)=z_0$, where $z_0$ is some constant, $AdS_4$ slice are located at $z_{\rm AdS,1}(t_h)=\sqrt{l^2+t_h^2}-\sqrt{l^2-1}$ (when $X_4>0$) and $z_{\rm AdS,2}(t_h)=\sqrt{l^2+t_h^2}+\sqrt{l^2-1}$ (when $X_4<0$) both sides of turn around point $X_4=0$($X_4$ being one of parametrization of $AdS_5$ defined in \cite{mismatched-branes}). At $t_h=0$, $z_{\rm AdS,min}=l \mp \sqrt{l^2-1}$. Minkowski and AdS brane can coexist for fixed value of $z$ beyond $z_{\rm AdS,min}$. Metric on $AdS_4$ brane is
\begin{eqnarray}
\label{metric-AdS4-KR}
ds^2=-d\tau_h^2+a(\tau_h) dH_3^2,
\end{eqnarray}
where $a(\tau_h)=\ \sin\left(\tau_h/l\right)$. $dS$ branes exist at  $z_{\rm dS,1}(t_h)=\sqrt{l^2+t_h^2}+\sqrt{l^2+1}$ and $z_{\rm dS,2}(t_h)=\sqrt{l^2+1}-\sqrt{l^2+t_h^2}$ with metric on each $dS_4$ brane
\begin{eqnarray}
\label{metric-dS4-KR}
ds^2=-d\tau_h^2+a(\tau_h) dH_3^2,
\end{eqnarray}
where $a(\tau_h)=\ \sinh\left(\tau_h/l\right)$.

{\bf Comment on the Wedge Holographic Realization of Mismatched Branes}: One can construct doubly holographic setup from (\ref{metric-AdS5-KR}) using the idea of AdS/BCFT. Let us state the three possible descriptions of doubly holographic setup constructed from (\ref{metric-AdS5-KR}).
\begin{itemize}
\item {\bf Boundary description}: $4D$ quantum field theory (QFT) at conformal boundary of (\ref{metric-AdS5-KR}).

\item {\bf Intermediate description}: Dynamical gravity localized on $4D$ end-of-the-world brane coupled to $4D$ boundary QFT.

\item {\bf Bulk description}: $4D$ QFT defined in the first description has $5D$ gravity dual whose metric is (\ref{metric-AdS5-KR}).
\end{itemize}

Due to covariant nature of AdS/CFT duality it remains the same if one works with the changed coordinates in the bulk i.e. different parametrisations of AdS does not imply different dualities\footnote{We thank K.~Skenderis to clarify this to us and pointing out his interesting paper \cite{Kostas}.} and therefore in the above doubly holographic setup, we expect defect to be $3$-dimensional conformal field theory because $4$-dimensional gravity is just FRW parametrization of $AdS_4$ spacetime (\ref{metric-AdS4-KR}). Relationship between boundary and bulk description is due to AdS/CFT correspondence, in particular, this kind of duality was studied in \cite{Kostas} where bulk is de-Sitter parametrization of $AdS_4$ and conformal field theory is QFT on $dS_3$. As discussed in detail in appendix {\bf A} of \cite{mismatched-branes} and summarised in this subsection that one can also have de-Sitter and Minkowski branes in this particular coordinate system (\ref{metric-AdS5-KR}). If one works with de-Sitter metric (\ref{metric-dS4-KR}) on end-of-the-world brane then we expect defect CFT to be non-unitary. Due to dynamical nature of gravity on Karch-Randall brane, holographic dictionary is not well understood in the braneworld scenario.\par

Now let us discuss what is the issue in describing wedge holography with ``mismatched branes''. Wedge holography has ``defect CFT'' which comes due to dynamical gravity on Karch-Randall branes. Suppose we have two Karch-Randall branes with different geometry, one of them is AdS brane and the other one is de-Sitter brane. Then due to AdS brane, defect CFT should be unitary and due to de-Sitter brane, defect CFT should be non-unitary. It seems that we have two different CFTs at the same defect. This situation will not change even one considers four branes or in general $2n$ branes. Hence, one may not be able to describe ``multiverse'' with mismatched branes from wedge holography. That was just an assumption. Common boundary of multiverses $M_1$ and $M_2$ (described in Fig. \ref{CP-dS+AdS-Multiverse}) can't be the same even when geometry is (\ref{metric-AdS5-KR}) due to ``time-dependent'' position of branes. All the AdS branes in $M_1$ can communicate with each other via transparent boundary conditions at the defect and similarly all the de-Sitter branes in $M_2$ are able to communicate with each other. But there is no communication between $M_1$ and $M_2$ even in (\ref{metric-AdS5-KR}).

Therefore we conclude that we can create multiverse of same branes(AdS or de-Sitter) but not the mixture of two. Hence issue of mismatched branes do not alter from wedge holography perspective too. Multiverse of AdS branes exists forever whereas multiverse of de-Sitter branes has finite lifetime\footnote{We thank A.~Karch for very helpful discussions on the existence of de-Sitter branes and issue of mismatched branes in wedge holography.}.

\section{Application to Information Paradox}
\label{AIP}
\begin{figure}
\begin{center}
\includegraphics[width=0.8\textwidth]{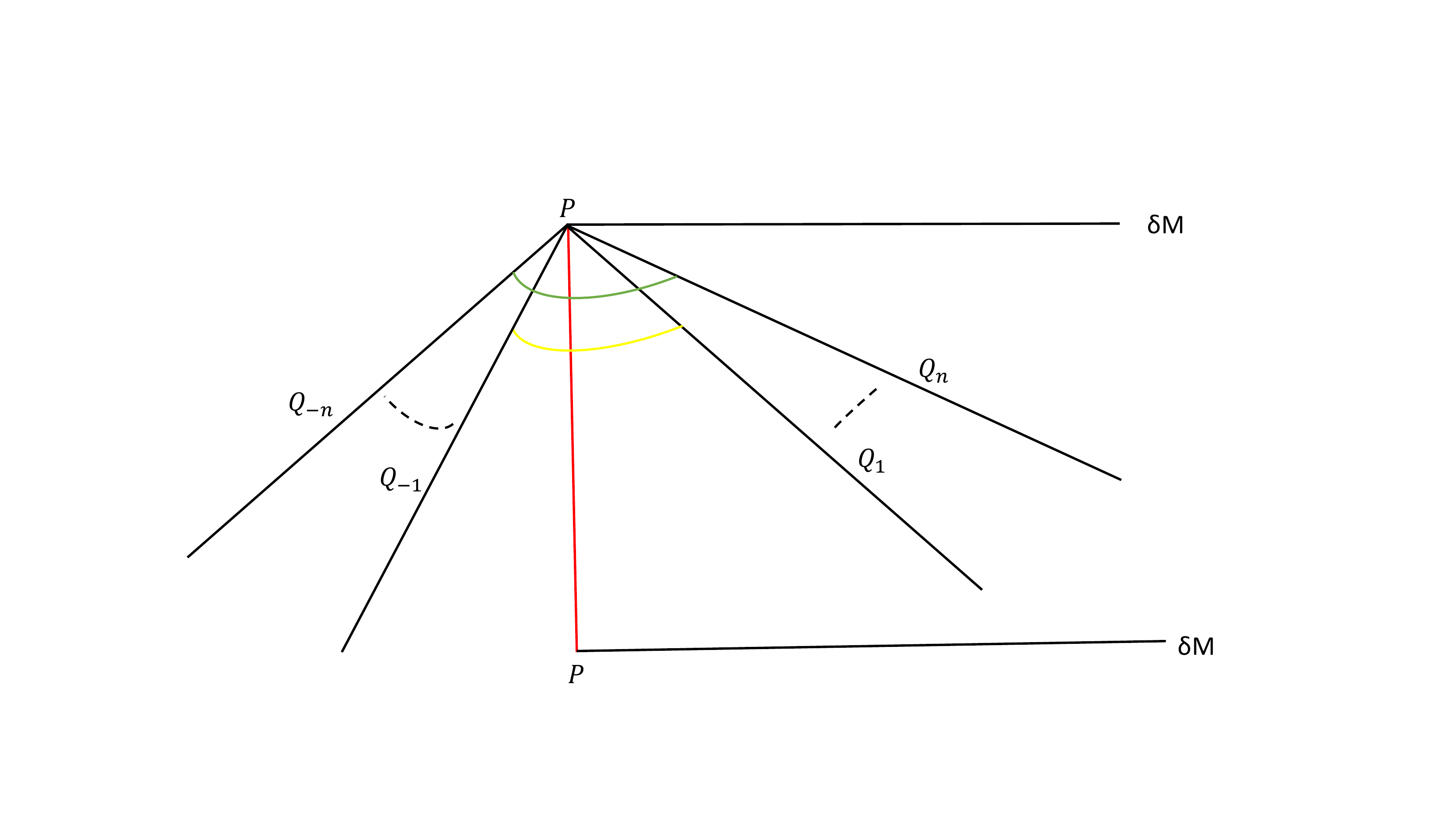}
\end{center}
\caption{In this figure, we assume that $n$ black holes contained in $Q_{1,2,...,n}$ emit Hawking radiation which is collected by baths $Q_{-1,-2,...,-n}$. Green and yellow curves represent island surfaces between $Q_{-n}$ and $Q_n$, $Q_{-1}$ and $Q_1$ respectively. The red curve represents the Hartman-Maldacena surface starting at the defect and meets its thermofield double partner. $\delta M$ is the AdS boundary.}
\label{BH+MBs}
\end{figure}

Multiverse consists of $2n$ Karch-Randall branes embedded in the bulk $AdS_{d+1}$. Therefore there will be a single Hartman-Maldacena surface connecting the defect CFTs between thermofield double partner and $n$ island surfaces (${\cal I}_1$,${\cal I}_2$,.....,${\cal I}_n$). $n$ Island surfaces will be stretching between corresponding branes of the same locations with opposite sign $(r=\pm n \rho)$, see Fig. \ref{BH+MBs}. Let us make the precise statement of a wedge holographic dictionary.

\fbox{\begin{minipage}{38em}{\it 
Classical gravity in $(d+1)$-dimensional AdS bulk\\ $\equiv$ (Quantum) gravity on $2n$ $d$-dimensional Karch-Randall branes with metric $AdS_d/dS_d$\\ $\equiv$ CFT living on $(d-1)$-dimensional defect.}
\end{minipage}}\\ \\
If the metric on Karch-Randall branes will be the de-Sitter metric then CFT will be non-unitary. Therefore this description is the same as the usual wedge holography with two Karch-Randall branes, the only difference is that we have $2n$ Karch-Randall branes now.
\par
Now let us write the explicit formula for entanglement entropies. We consider $Q_{1,2,....,n}$ as black holes which emit Hawking radiation, the radiation is collected by gravitating baths $Q_{-1,-2,....,-n}$ (Fig. \ref{BH+MBs}). In this setup, entanglement entropy for the islands surfaces will be:
\begin{eqnarray}
\label{Island-Formula-n-BHs}
& & S_{\rm Island}=S_{Q_{-1}-Q_1}^{{\cal I}_1} + S_{Q_{-2}-Q_2}^{{\cal I}_2}+.......+S_{Q_{-n}-Q_n}^{{\cal I}_{n}}.
\end{eqnarray}

If entanglement entropy corresponding to the Hartman-Maldacena surface, i.e.,$S_{\rm HM} \propto t$ and $S_{\rm Island}=2 S^{{i=1,2,..,n}, \ \rm thermal}_{\rm BH}$ then we can get the Page curve, where $S_{\rm Island}$ and $S_{\rm HM}$ can be calculated using Ryu-Takayanagi formula \cite{RT}.
Following are the three descriptions of the multiverse: 
\begin{itemize}
\item {\bf Boundary Description:} BCFT is living at the $AdS_{d+1}$ boundary with $(d-1)$-dimensional boundary.
\item {\bf Intermediate Description:} $2n$ gravitating systems interact with each other via transparent boundary conditions at the  $(d-1)$-dimensional defect.
\item {\bf Bulk Description:} Gravity dual of BCFT is Einstein gravity in the bulk.
\end{itemize} 

{\bf Consistency Check:}
Let us check the formula given in (\ref{Island-Formula-n-BHs}) for $n=2$. 
\subsection{Page Curve of Eternal AdS Black Holes in $n=2$ Multiverse}
\label{PC-TEBH}
First, we will calculate the thermal entropies of black holes. The metric of the black holes in $AdS$ background is:
\begin{eqnarray}
\label{metric-bulk-BH}
ds_{(d+1)}^2=g_{\mu \nu} dx^\mu dx^\nu=dr^2+\cosh^2(r)\Biggl(\frac{\frac{dz^2}{f(z)}-f(z) dt^2+\sum_{i=1}^{d-2}dy_i^2 }{z^2}\Biggr),
\end{eqnarray}
 where $f(z)=1-\frac{z^{d-1}}{z_h^{d-1}}$. For $z=z_h$, thermal entropy has the following form(we set $z_h=1$ throughout the calculation for the simplicity and focus on $d=4$):
\begin{eqnarray}
\label{thermal}
S_{\rm AdS}^{\rm thermal} = \frac{A^{\rm BH}_{z=z_h}}{4 G_N^{(d+1)}}= \frac{1}{4 G_N^{(5)}}\int dr \cosh^{2}(r) \int dy_1 \int dy_2 = \frac{V_{2}}{4 G_N^{(5)}} \int dr \cosh^{2}(r),
\end{eqnarray} 
where $V_2=\int \int dy_1 dy_2$. 
Let's consider the $n=2$ case, in which two Karch-Randall branes between $-2 \rho \leq r \leq 2 \rho$ and $ -\rho \leq r \leq \rho$ act as a black hole and bath systems. Therefore total thermal entropies for two eternal AdS black holes will be:
\begin{eqnarray}
\label{s-thermal-total-AdS}
& & \hskip -0.9in
S_{\rm AdS}^{\rm thermal, \ total} = \frac{V_{2}}{4 G_N^{(5)}} \int_{-2 \rho}^{2 \rho} dr \cosh^{2}(r)+ \frac{V_{2}}{4 G_N^{(5)}} \int_{-\rho}^{ \rho} dr \cosh^2(r)  \nonumber\\
& &=\frac{V_{2}}{4 G_N^{(5)}}\left(\frac{1}{2} (6 \rho +\sinh (2 \rho )+\sinh (4 \rho ))\right).
\end{eqnarray}
Now let us obtain the Page curve using the formula given in (\ref{Island-Formula-n-BHs}) for two eternal black holes.

{\bf Entanglement entropy contribution from Hartman-Maldacena surface}: Bulk metric (\ref{metric-bulk-BH}) in terms of infalling Eddington-Finkelstein coordinate, $d v= dt -\frac{dz}{f(z)}$ simplified as follows.
\begin{eqnarray}
\label{metric-bulk-BH-EFC}
ds_{(4+1)}^2=g_{\mu \nu} dx^\mu dx^\nu=dr^2+\cosh^2(r)\Biggl(\frac{-f(z) dv^2-2 dv dz+\sum_{i=1}^{2}dy_i^2 }{z^2}\Biggr).
\end{eqnarray}
Induced metric for the Hartman-Maldacena surface parametrize by $r \equiv r(z)$ and $v \equiv v(z)$ obtained as:
\begin{eqnarray}
\label{induced-metric-HM}
& & ds^2= \Biggl({r'(z)^2-\frac{\cosh^2(r(z))v'(z)}{z^2} \left(2+f(z)v'(z)\right)}\Biggr) dz^2+ \frac{\cosh^2(r(z))}{z^2}\sum_{i=1}^{2}dy_i^2,
\end{eqnarray}

where $r'(z)=\frac{dr}{dz}$ and $v'(z)=\frac{dv}{dz}$. From (\ref{induced-metric-HM}), the area of the Hartman-Maldacena surface is obtained as:
\begin{eqnarray}
& & A_{\rm HM}^{\rm AdS}=V_{2} \int_{z_1}^{z_{\rm max}} dz \Biggl( \frac{\cosh^2(r(z))}{z^2} \sqrt{ {r'(z)^2-\frac{\cosh^2(r(z))v'(z)}{z^2} \left(2+f(z)v'(z)\right)}}\Biggr),
\end{eqnarray}
where $z_1$ is the point on gravitating bath, $z_{\rm max}$ is the turning point of Hartman-Maldacena surface and $V_{2}=\int \int dy_1 dy_2$. For large time, i.e., $t\rightarrow \infty$, $r(z) \rightarrow 0$ \cite{Massless-Gravity}. Therefore,
\begin{eqnarray}
\label{AHM-AdS}
& & A_{\rm HM}^{\rm AdS}=V_{2} \int_{z_1}^{z_{\rm max}} dz\Biggl(   \frac{\sqrt{- v'(z)\left(2+f(z)v'(z)\right)}}{z^3}\Biggr).
\end{eqnarray}
Equation of motion for the embedding $v(z)$ is
\begin{eqnarray}
& & 
\frac{d}{dz}\left(\frac{\partial L}{\partial v'(z)}\right)=0, \nonumber\\
& & \implies \frac{\partial L}{\partial v'(z)}=E, \nonumber\\
& &\implies v'(z)=\frac{-E^2 z^6-\sqrt{E^4 z^{12}+E^2 f(z) z^6}-f(z)}{E^2 f(z) z^6+f(z)^2}.
\end{eqnarray}
Since, $ v'(z)|_{z=z_{\rm max}} =0$ where $z_{\rm max}$ is the turning point therefore, $E=\frac{i\sqrt{f(z_{\rm max})}}{z_{\rm max}^3}$ and $\frac{dE}{dz_{\rm max}}=0$ implies $z_{\rm max}=\frac{7 z_h}{6}$ (i.e. turning point of Hartman-Maldacena surface is outside the horizon). We can obtain time on the bath as given below:
\begin{eqnarray}
\label{tz1}
t_1=t(z_1)=-\int_{z_1}^{z_{\rm max}} \left(v'(z)+\frac{1}{f(z)}\right)dz.
\end{eqnarray}
Now let us analyze, the late-time behavior of the area of the Hartman-Maldacena surface:
\begin{eqnarray}
& & {\rm lim}_{t \rightarrow \infty} \frac{dA_{\rm HM}^{\ \rm AdS}}{dt}={\rm lim}_{t \rightarrow \infty}\Biggl( \frac{\frac{dA_{\rm HM}^{\rm AdS}}{dz_{\rm max} }}{\frac{dt}{dz_{\rm max} }}\Biggr) = \frac{L(z_{\rm max},v'(z_{\rm max}))+\int_{z_1}^{z_{\rm max}} \frac{\partial L}{\partial z_{\rm max} }dz}{-v'(z_{\rm max})-\frac{1}{f(z_{\rm max})}-\int_{z_1}^{z_{\rm max}}\frac{\partial v'(z)}{\partial z_{\rm max} }}.
\end{eqnarray}
Since,
\begin{eqnarray}
& & 
{\rm lim}_{t \rightarrow \infty}\frac{\partial v'(z)}{\partial z_{\rm max} }={\rm lim}_{t \rightarrow \infty}\frac{\partial v'(z)}{\partial E}\frac{\partial E}{\partial z_{\rm max} }=0, \nonumber\\
& & {\rm lim}_{t \rightarrow \infty}\frac{\partial L(z,v'(z))}{\partial z_{\rm max} } = \frac{\partial L(z,v'(z))}{\partial v'(z) } \frac{\partial v'(z)}{\partial z_{\rm max} }=0.
\end{eqnarray}
Therefore,
\begin{eqnarray}
{\rm lim}_{t \rightarrow \infty} \frac{dA_{\rm HM}^{\rm AdS}}{dt} = \frac{L(z_{\rm max},v'(z_{\rm max}))}{-v'(z_{\rm max})-\frac{1}{f(z_{\rm max})}}  = \frac{\frac{\sqrt{-v'(z_{\rm max})(2+f(z_{\rm max})v'(z_{\rm max}))}}{z_{\rm max}^3}}{-v'(z_{\rm max})-\frac{1}{f(z_{\rm max})}} = constant.
\end{eqnarray}

The above equation implies that $A_{\rm HM}^{\rm AdS} \propto t_1$, and hence entanglement entropy for the Hartman-Maldacena surface has the following form
\begin{eqnarray}
\label{SHM-AdS}
& & S_{\rm HM}^{\rm AdS} \propto t_1.
\end{eqnarray}
This corresponds to an infinite amount of Hawking radiation when $t_1 \rightarrow \infty$, i.e., at late times, and hence leads to information paradox.

{\bf Entanglement entropy contribution from Island surfaces}: Now consider the island surfaces parametrize as $t=constant$ and $z \equiv z(r)$. Entanglement entropy of two eternal AdS black holes for the island surfaces can be obtained using (\ref{Island-Formula-n-BHs}). Since there are two island surfaces(${\cal I}_1$ and ${\cal I}_2$) stretching between the Karch-Randall branes located at $r=\pm \rho$ (${\cal I}_1$) and $r=\pm 2 \rho$ (${\cal I}_2$), and hence we can write (\ref{Island-Formula-n-BHs}) for the same as given below
\begin{eqnarray}
\label{EE-Island}
& & 
S_{\rm AdS}^{\rm Island} =S_{Q_{-1}-Q_1}^{{\cal I}_1} + S_{Q_{-2}-Q_2}^{{\cal I}_2} =  \frac{\left({\cal A}_{{\cal I}_1}+{\cal A}_{{\cal I}_2}\right)}{4 G_N^{(5)}}=
\frac{ \int d^3x \sqrt{h_1}+\int d^3x \sqrt{h_2}}{4 G_N^{(5)}}.
\end{eqnarray}
First we calculate ${\cal A}_{{\cal I}_1}$. Induced metric on Karch-Randall branes can be obtained from (\ref{metric-bulk-BH}) by using the parametrization of island surface as $t =constant$ and $z=z(r)$ and restricting to $d=4$ with $f(z)=1-z^3$ (since $z_h=1$), 
\begin{eqnarray}
\label{induced-metric-AdS-IS}
& & ds^2= \Biggl({1+\frac{\cosh^2(r)z'(r)^2}{z(r)^2(1-z(r)^{3})}}\Biggr) dr^2+ \frac{\cosh^2(r)}{z(r)^2}\sum_{i=1}^{2}dy_i^2,
\end{eqnarray}

Area of the island surface ${\cal I}_1$ from (\ref{induced-metric-AdS-IS}) obtained as
\begin{eqnarray}
\label{AIS-AdS}
{\cal A}_{{\cal I}_1}=V_{2}\int_{- \rho}^{\rho} dr {\cal L}_{{\cal I}_1}\left(z(r),z'(r)\right) =V_{2}\int_{- \rho}^{\rho} dr\Biggl(\frac{\cosh^{2}(r)}{z(r)^{2}}\sqrt{1+\frac{\cosh^2(r)z'(r)^2}{z(r)^2(1-z(r)^{3})}}\Biggr),
\end{eqnarray} 
where we have chosen $z_h=1$ and hence $0<z<1$ for $f(z)\geq 0$.
Let us discuss the variation of the action (\ref{AIS-AdS}).
\begin{eqnarray}
\label{var-AIS}
& & 
\delta {\cal A}_{{\cal I}_1} = V_{2}\int_{- \rho}^{\rho} dr \Biggl[\left(\frac{\delta{\cal L}_{{\cal I}_1}\left(z(r),z'(r)\right)}{\delta z(r)}\right) \delta z(r)+\left(\frac{\delta{\cal L}_{{\cal I}_1}\left(z(r),z'(r)\right)}{\delta z'(r)}\right) \delta z'(r) \Biggr]\nonumber\\
& &\hskip-0.2in = V_{2}\int_{- \rho}^{\rho} dr \left(\frac{\delta{\cal L}_{{\cal I}_1}\left(z(r),z'(r)\right)}{\delta z'(r)}\right) \delta z(r)- \int_{- \rho}^{\rho} dr \Biggl[\frac{d}{dr}\left(\frac{\delta{\cal L}_{{\cal I}_1}\left(z(r),z'(r)\right)}{\delta z'(r)}\right)-\left(\frac{\delta{\cal L}_{{\cal I}_1}\left(z(r),z'(r)\right)}{\delta z(r)}\right) \Biggr]\delta z(r). \nonumber\\
\end{eqnarray}
Variational principle will be meaningful only if first term of the above equation vanishes. Second term is the EOM for the embedding $z(r)$. Let us see what this implies
\begin{eqnarray}
\label{VP-meaningful}
& & \int_{- \rho}^{\rho} dr \left(\frac{\delta{\cal L}_{{\cal I}_1}\left(z(r),z'(r)\right)}{\delta z'(r)}\right) \delta z(r) = \int_{- \rho}^{\rho} dr \Biggl(\frac{\cosh ^4(r) z'(r)}{z(r)^4 f(z(r)) \sqrt{\frac{\cosh ^2(r) z'(r)^2}{z(r)^2 f(z(r))}+1}} \Biggr)\delta z(r),
\end{eqnarray}
(\ref{VP-meaningful}) vanishes either we impose Dirichlet boundary condition on the branes, i.e., $\delta z(r=\pm \rho)=0$ or Neumann boundary condition on the branes, i.e., $z'(r=\pm \rho)=0$. For gravitating baths Neumann boundary condition allow RT surfaces to move along the branes. In this case, minimal surface is the black hole horizon \cite{GB-3}.

\begin{figure}
\begin{center}
\includegraphics[width=0.7\textwidth]{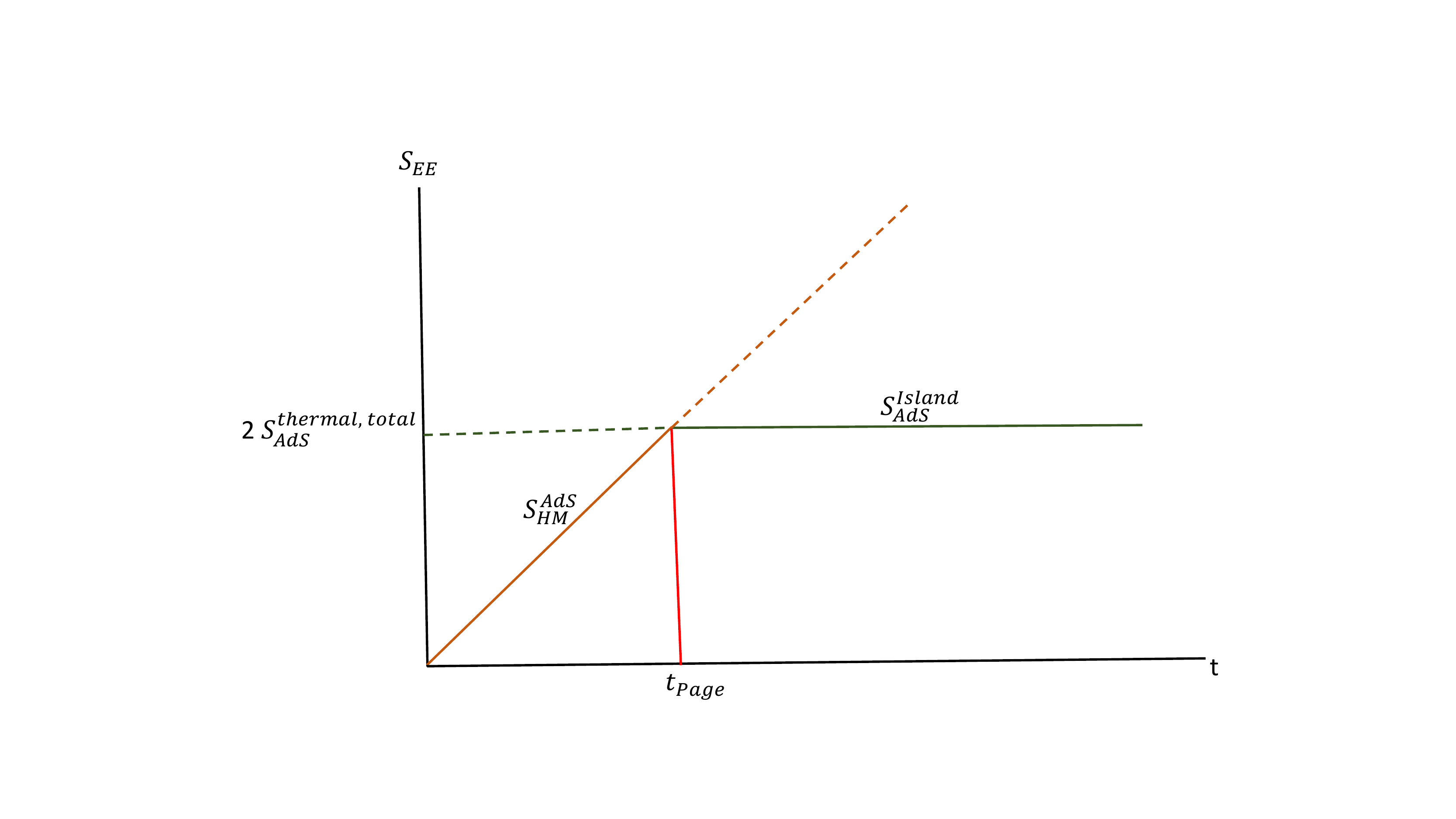}
\end{center}
\caption{Page curve of eternal AdS black holes for $n=2$ multiverse.}
\label{PC-AdS}
\end{figure}

 Euler-Lagrange equation of motion for the embedding $z(r)$ from the action (\ref{AIS-AdS}) turns out to be:
{\footnotesize
\begin{eqnarray}
\label{EOM}
& & \hskip -0.3in
\frac{\cosh ^2(r) }{2 z(r)^4 \left(z(r)^3-1\right) \left(-\cosh ^2(r) z'(r)^2+z(r)^5-z(r)^2\right) \sqrt{\frac{\cosh ^2(r)
   z'(r)^2}{z(r)^2-z(r)^5}+1}} \nonumber\\
   & & \hskip -0.5in \times \Biggl(z(r)^4 \cosh ^2(r) z'(r)^2+2 z(r) \cosh ^2(r) z'(r)^2+6 \sinh (r) \cosh ^3(r) z'(r)^3
   -2 z(r)^5 \cosh (r)
   \left(\cosh (r) z''(r)  +4 \sinh (r) z'(r)\right)\nonumber\\
   & &+2 z(r)^2 \cosh (r) \left(\cosh (r) z''(r)+4 \sinh (r) z'(r)\right)+4 z(r)^9-8
   z(r)^6+4 z(r)^3\Biggr)
   =0.
\end{eqnarray}
}
Interestingly, solution of (\ref{EOM}) is $z(r)=1$ which is black hole horizon and its satisfies Neumann boundary condition on the branes. {The same can be seen from the structure of (\ref{EOM}). Terms inside the open bracket of (\ref{EOM}) contains mostly $z'(r)$ and $z''(r)$, but there is a particular combination independent of $z'(r)$ and $z''(r)$, $(4 z(r)^9-8 z(r)^6+4 z(r)^3)$ which vanishes for $z(r)=1$ and hence $z(r)=1$ is the solution of (\ref{EOM}).} This implies Ryu-Takayanagi surface is the black hole horizon because $z_h=1$\footnote{It was discussed in \cite{GB-3} that Neumann boundary condition on gravitating branes implies that Ryu-Takayanagi surface in the wedge holography is the black hole horizon. The same was also obtained in \cite{Massless-Gravity} by using inequality condition on the area of island surface. We obtained the same throughout the paper wherever we have discussed the entanglement entropy of island surfaces.} and on substituting $z(r)=1$ in (\ref{AIS-AdS}), we obtained the minimal area of the island surface ${\cal I}_{1}$ as
\begin{eqnarray}
\label{AIS-AdS-i}
{\cal A}_{{\cal I}_1}=V_{2}\int_{- \rho}^{\rho} dr {\cosh^{2}(r)}.
\end{eqnarray}
Minimal area of the second island surface ${\cal I}_{2}$ will be the same as (\ref{AIS-AdS-i}) with different limits of integration due to different locations of Karch-Randall branes ($r=\pm 2 \rho$).
\begin{eqnarray}
\label{AIS-AdS-ii}
{\cal A}_{{\cal I}_2}=V_{2}\int_{-2 \rho}^{2 \rho} dr {\cosh^{2}(r)}.
\end{eqnarray}
Substituting (\ref{AIS-AdS-i}) and (\ref{AIS-AdS-ii}) into (\ref{EE-Island}), we obtain the total entanglement entropy of island surfaces
\begin{eqnarray}
\label{EE-Island-simp}
& & 
S_{\rm AdS}^{\rm Island}  =  \frac{2 V_{2}}{4 G_N^{(5)}}\Biggl(\int_{- \rho}^{\rho} dr {\cosh^{2}(r)}+\int_{-2 \rho}^{2 \rho} dr {\cosh^{2}(r)}\Biggr)=2 S_{\rm AdS}^{\rm thermal, \ total}.
\end{eqnarray}
prefactor $2$ in (\ref{EE-Island-simp}) comes due to the extra two island surfaces from the thermofield double partner. From (\ref{SHM-AdS}) and (\ref{EE-Island-simp}), we obtain the Page curve for $n=2$ multiverse as shown in Fig. \ref{PC-AdS}.

\subsection{Page Curve of Schwarzschild de-Sitter Black Hole}
\label{IP-SdS} 
\begin{figure}
\begin{center}
\includegraphics[width=0.6\textwidth]{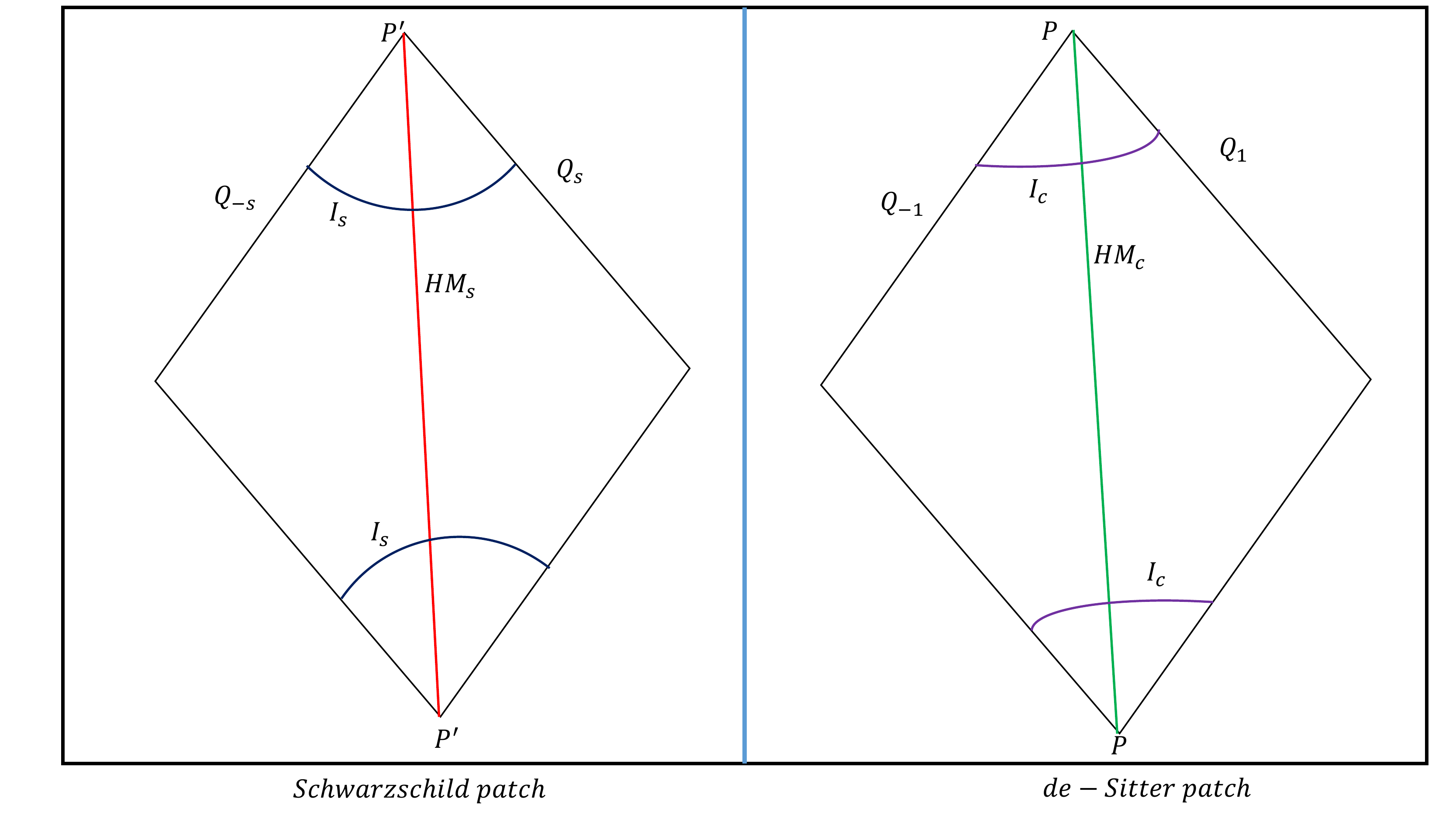}
\end{center}
\caption{Realization of Schwarzschild de-Sitter black hole in wedge holography. $I_s$ and $I_c$ are black hole and cosmological island surfaces (black hole and de-Sitter horizons in our case). Red ($HM_s$) and green ($HM_c$) lines are Hartman-Maldacena surfaces for Schwarzschild and de-Sitter patches. $Q_s$ and $Q_1$ branes consist of Schwarzschild and de-Sitter patches. $Q_{-s}$ and $Q_{-1}$ are baths to collect Hawking and Gibbons-Hawking radiation emitted by black hole and cosmological horizons.}
\label{SdS-CP}
\end{figure}

In this section, we study the information paradox of the Schwarzschild de-Sitter black hole. As discussed in section \ref{AdS+dS-Multiverse}, we can not have mismatched branes connected at the same defect. Therefore, we study this problem in two parts by first calculating the Page curve of the Schwarzschild patch and then the Page curve of the de-Sitter patch similar to the non-holographic model \cite{Gopal+Nitin}. This can be done as follows. We study the Schwarzschild patch in subsection \ref{Sch-patch} where we consider two flat space branes embedded in the bulk and de-Sitter patch in subsection \ref{dS-patch} with two de-Sitter branes. We have shown the setup in Fig. \ref{SdS-CP}. The setup is two copies of wedge holography with flat space and de-Sitter branes in Schwarzschild and de-Sitter patches respectively.

\subsubsection{Schwarzschild patch} 
\label{Sch-patch}
Since for Schwarzschild black hole, $\Lambda=0$, therefore to realize Schwarzschild black hole on Karch-Randall brane, we need to consider the flat space black holes. It was shown in \cite{WH-2} that one can get flat space black holes on Karch-Randall branes provided bulk metric should have the following form:
\begin{equation}
\label{metric-bulk-flat-space}
ds_{(d+1)}^2=g_{\mu \nu} dx^\mu dx^\nu=dr^2+e^{2 r} h_{ij} dy^i dy^j =dr^2+{e^{2 r}} \Biggl(\frac{\frac{dz^2}{f(z)}-f(z) dt^2+\sum_{i=1}^{d-2}dy_i^2 }{z^2}\Biggr).
\end{equation}
Induced metric $h_{ij}$ on the brane given in (\ref{metric-bulk-flat-space}) obey the following Einstein equation on the brane
\begin{eqnarray}
\label{Brane-Einstein-equation-vacuum}
R_{ij}-\frac{1}{2}h_{ij} R[h_{ij}] =0.
\end{eqnarray} 
(\ref{Brane-Einstein-equation-vacuum}) is the equation of motion of the following Einstein-Hilbert term on the brane:
\begin{eqnarray}
\label{boundary-EH-action-FS}
S_{\rm FS}^{\rm EH} =\lambda^{\rm FS} \int d^{d}x \sqrt{-h} R[h],
\end{eqnarray}
where $\lambda^{\rm FS}\left(\equiv \frac{1}{16 \pi G_N^{d}}=\frac{1}{16 \pi G_N^{(d+1)}}\frac{e^{(d-2)a_1}}{(d-2)}\right)$ encodes information about the effective Newton's constant in $d$ dimensions and (\ref{boundary-EH-action-FS}) has been obtained from substitution of  (\ref{metric-bulk-flat-space}) into the (\ref{bulk-action}). For Schwarzschild black hole in $d$-dimensions $f(z)=1-\frac{z_h^{d-3}}{z^{d-3}}$ \cite{Hoshimoto et al}. Further, metric (\ref{metric-bulk-flat-space}) satisfy Neumann boundary condition at $r=constant$  with brane tension $T_{\rm flat \ space}= |d-1|$. Schwarzchild black hole and its bath will be given by two Karch-Randall branes located at $r=\pm a_1$. Thermal entropy of Schwarzschild patch can be obtained from (\ref{metric-bulk-flat-space}) for $z=z_h$ and the final result is given as 
\begin{eqnarray}
\label{TE-Sch}
S_{\rm thermal}^{\rm Schwarzschild}= \frac{ V_2 \int_{-a_1}^{a_1} dr e^{2r}}{4 G_N^{(5)}}=\frac{ V_2 \sinh (2 a_1)}{4 G_N^{(5)}}.
\end{eqnarray}

{\bf Hartman-Maldacena Surface}: Defining infalling Eddington-Finkelstein coordinate: $d v= dt -\frac{dz}{f(z)}$, flat space metric (\ref{metric-bulk-flat-space}) simplifies to:
\begin{eqnarray}
\label{metric-FS-branes}
& & ds^2 = dr^2+ \frac{e^{2r}}{z^2} \left(-f(z)dv^2-2 dv dz+\sum_{i=1}^2 dy_i^2\right).
\end{eqnarray}
Induced metric for the Hartman-Maldacena surface parametrized by $r=r(z)$ and $v=v(z)$ is
\begin{eqnarray}
\label{induced-metric-HM-FS}
& &ds^2= \Biggl({r'(z)^2-\frac{e^{2r(z)}}{z^2} \left(2+f(z)v'(z)\right)}\Biggr) dz^2+ \frac{e^{2r(z)}}{z^2}\sum_{i=1}^{2}dy_i^2.
\end{eqnarray}
Area of the Hartman-Maldacena surface using (\ref{induced-metric-HM-FS}) obtained as:
\begin{eqnarray}
\label{AHM-Sch}
& & A_{\rm HM}^{\rm Schwarzschild}=V_2 \int_{z_1}^{z_{\rm max}} dz  \Biggl(\frac{e^{2r(z)}}{z^2} \sqrt{r'(z)^2-\frac{e^{2r(z)} v'(z)}{z^2}\left(2+f(z)v'(z)\right)}\Biggr).
\end{eqnarray}
For large time, i.e., $t\rightarrow \infty$, $r(z) \rightarrow 0$\footnote{We can show the same by following the steps given in detail from (\ref{AHM-de-Sitter})-(\ref{soln-r(z)}). But we have to replace the warp factor $\sinh(r(z))$ by $e^{r(z)}$.} \cite{Massless-Gravity}. Therefore,
\begin{eqnarray}
\label{AHM-Sch-i}
& & A_{\rm HM}^{\rm Schwarzschild}=V_2 \int_{z_1}^{z_{\rm max}} dz   \Biggl(\frac{\sqrt{- v'(z)\left(2+f(z)v'(z)\right)}}{z^3}\Biggr).
\end{eqnarray}
Since the area of the Hartman-Maldacena surface is similar to (\ref{AHM-AdS}) except the volume factor, here we are restricted to $d=4$, therefore for the Schwarzschild patch too, $A_{\rm HM}^{\rm Schwarzschild} \propto t_1$. Therefore entanglement entropy contribution from the Hartman-Maldacena surface of the Schwarzschild patch has the linear time dependence
\begin{eqnarray}
\label{SHM-t}
S_{\rm HM}^{\rm Schwarzschild} \propto t_1.
\end{eqnarray}

\begin{figure}
\begin{center}
\includegraphics[width=0.7\textwidth]{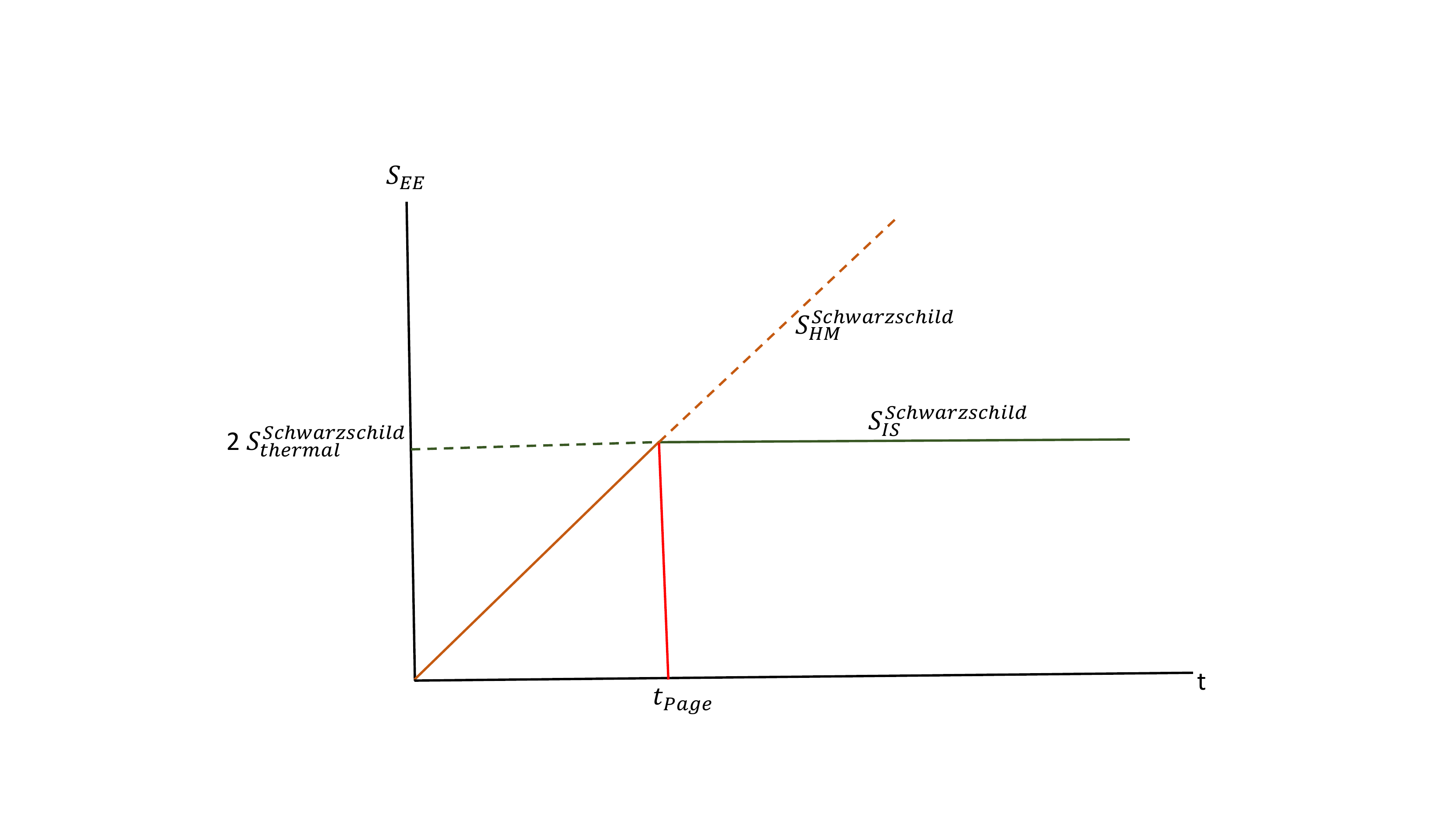}
\end{center}
\caption{Page curve of Schwarzschild patch.}
\label{PC-Sch}
\end{figure}

{\bf Island Surface}: The island surface is parametrized by $t=constant$ and $z=z(r)$. The area of island surface can be obtained from the induced metric in terms of embedding($z(r)$) and its derivative using the bulk metric (\ref{metric-bulk-flat-space}), induced metric is
\begin{eqnarray}
\label{induced-metric-Sch-IS}
& & ds^2= \Biggl({1+\frac{e^{2r}z'(r)^2}{z(r)^2\left(1-\frac{1}{z(r)}\right)}}\Biggr) dr^2+ \frac{e^{2r}}{z(r)^2}\sum_{i=1}^{2}dy_i^2.
\end{eqnarray}
where we have used $f(z)=\left(1-\frac{1}{z}\right)$. Using (\ref{induced-metric-Sch-IS}) area of the island surface for the Schwarzschild patch is obtained as
\begin{eqnarray}
\label{AIS}
A_{\rm IS}^{\rm Schwarzschild}= V_2 \int_{-a_1}^{a_1} dr  \Biggl(\frac{e^{2r}}{z(r)^2}\sqrt{1+\frac{e^{2r}z'(r)^2}{z(r)^2\left(1-\frac{1}{z(r)}\right)}}\Biggr).
\end{eqnarray} 
In the above equation, we have set $z_h=1$ for simplicity and hence $f(z)\geq0$ requires $z>1$.
Substituting the Lagrangian of (\ref{AIS}) in (\ref{var-AIS}), first term of the last line of (\ref{var-AIS}) for (\ref{AIS}) implies
\begin{eqnarray}
\label{NBC-Sch}
\frac{e^{4 r} z'(r)}{\left(1-\frac{1}{z(r)}\right) z(r)^4 \sqrt{\frac{e^{2 r} z'(r)^2}{\left(1-\frac{1}{z(r)}\right) z(r)^2}+1}}=0. 
\end{eqnarray}
Therefore, we have well-defined variational principle of (\ref{AIS}) provided embedding function satisfies Neumann boundary condition on the branes, i.e., $z(r=\pm a_1)=0$ and hence the minimal surface will be the black hole horizon, i.e., $z(r)=1$ similar to \cite{GB-3,Massless-Gravity}. The same can be obtained from the equation of motion of $z(r)$ worked out as follows
{
\begin{eqnarray}
\label{EOM-Sch}
& &
\frac{e^{2 r} \sqrt{\frac{e^{2 r} z'(r)^2+z(r)^2-z(r)}{(z(r)-1) z(r)}} }{2 z(r)^2 \left(e^{2 r} z'(r)^2+z(r)^2-z(r)\right)^2} \Biggl(3 e^{2 r} z'(r)^2 \left(2 e^{2 r} z'(r)-1\right)
+2 z(r)^2 \left(e^{2 r} z''(r)+4 e^{2 r} z'(r)-4\right)\nonumber\\
& & 
+2 z(r) \left(-e^{2 r}
   z''(r)+e^{2 r} z'(r)^2-4 e^{2 r} z'(r)+2\right)+4 z(r)^3\Biggr)
=0.
\end{eqnarray}
}
Solution of (\ref{EOM-Sch}) is the black hole horizon, i.e. $z(r)=1$\footnote{See the terms inside the open bracket of (\ref{EOM-Sch}), there are terms with derivatives of $z(r)$ and a particular combination $(-8z(r)^2+4z(r)+4z(r)^3)$ which vanishes for $z(r)=1$.} consistent with the Neumann boundary condition on the branes \cite{GB-3}. Therefore the minimal area of the island surface can be obtained after substituting $z(r)=1$ in (\ref{AIS}) and the final result is:
\begin{eqnarray}
\label{AIS-simp-Schwarzschild}
A_{\rm IS}^{\rm Schwarzschild}= V_2 \int_{-a_1}^{a_1} dr e^{2r}=V_2 \sinh(2 a_1).
\end{eqnarray}
Therefore entanglement entropy for the island surface of the Schwarzschild patch is
\begin{eqnarray}
\label{SIS-Sch}
S_{\rm IS}^{\rm Schwarzschild}=\frac{A_{\rm IS}^{\rm Schwarzschild}}{4 G_N^{(5)}}= \frac{2 V_2 \int_{-a_1}^{a_1} dr e^{2r}}{4 G_N^{(5)}} =\frac{2 V_2 \sinh(2 a_1)}{4 G_N^{(5)}}
= 2 S_{\rm thermal}^{\rm Schwarzschild}.
\end{eqnarray}
Numerical factor $2$ in the above equation appear because of second island surface in thermofield double partner (see Fig. \ref{SdS-CP}). Therefore we can get the Page curve by plotting (\ref{SHM-t}) and (\ref{SIS-Sch}) for the Schwarzschild patch shown in Fig. \ref{PC-Sch}.

\subsubsection{de-Sitter patch}
\label{dS-patch}
 de-Sitter black hole and its bath will be located at $r=\pm \rho$. The Metric for the bulk which contains de-Sitter branes is 
\begin{eqnarray}
\label{de-Sitter-metric-SdS}
& & \hskip -0.3in
ds_{(d+1)}^2=g_{\mu \nu} dx^\mu dx^\nu=dr^2+\sinh^2(r) h_{ij}^{\rm dS} dy^i dy^j =dr^2+\sinh^2(r) \Biggl(\frac{\frac{dz^2}{f(z)}-f(z) dt^2+\sum_{i=1}^{d-2}dy_i^2 }{z^2}\Biggr),\nonumber\\
\end{eqnarray}
where in $d=4$, for de-Sitter space: $f(z)=1-\frac{\Lambda}{3}z^2=1-\left(\frac{z}{z_s}\right)^2$ where $z_s=\sqrt{\frac{3}{\Lambda}}$. Thermal entropy of the de-Sitter patch can be obtained from (\ref{de-Sitter-metric-SdS}) by setting $z_s=1$\footnote{We used $z_s=1$ only for the simplification of calculation. Since cosmological constant is very small and hence in reality $z_s>>1$ but some number which will not affect our qualitative results.} in the same and the result is
\begin{eqnarray}
\label{thermal-dS}
S_{\rm dS}^{\rm thermal}=\frac{ A_{z=z_s}}{4 G_N^{(5)}} = \frac{ V_2 \int_{-\rho}^{\rho} dr \sinh^{2}(r)}{4 G_N^{(5)}}=\frac{ V_2 \left(\sinh (\rho ) \cosh (\rho )-\rho\right)}{4 G_N^{(5)}},
\end{eqnarray}
where $V_2=\int \int dy_1 dy_2$.\\
{\bf Hartman-Maldacena Surface}:
Similar to Schwarzschild patch, we define: $d v= dt -\frac{dz}{f(z)}$, and hence (\ref{de-Sitter-metric-SdS}) becomes:
\begin{eqnarray}
\label{metric-dS-EFK}
& & ds^2 = dr^2+ \sinh^{2}(r) \left(\frac{-f(z)dv^2-2 dv dz+\sum_{i=1}^2 dy_i^2}{z^2}\right).
\end{eqnarray}
Parametrization of Hartman-Maldacena surface is $r=r(z)$ and $v=v(z)$ and hence the area of the same can be obtained using (\ref{metric-dS-EFK}) for the aforementioned parametrization and written below:
\begin{eqnarray}
\label{AHM-de-Sitter}
& & \hskip -0.3in A_{\rm HM}^{\rm de-Sitter}=V_2 \int_{z_1^{\rm dS}}^{z_{\rm max}^{\rm dS}} dz {\cal L}_{\rm HM}^{\rm dS}
=V_2 \int_{z_1^{\rm dS}}^{z_{\rm max}^{\rm dS}} dz  \Biggl(\frac{\sinh^{2}(r(z))}{z^2}  \sqrt{r'(z)^2-\frac{\sinh^{2}(r(z))  v'(z)}{z^2}\left(2+f(z)v'(z)\right)}\Biggr),\nonumber\\
\end{eqnarray}
where $z_1^{\rm dS}$ and $z_{\rm max}^{\rm dS}$ are the point on gravitating bath and turning point of Hartman-Maldancena surface for the de-Sitter geometry. In (\ref{AHM-de-Sitter}), $v(z)$ is cyclic therefore conjugate momentum of $v(z)$ is constant, i.e., $\frac{\partial {\cal L}_{\rm HM}^{\rm dS}}{\partial v'(z) }=C$ ($C$ being the constant) implies
{\footnotesize
\begin{eqnarray}
\label{v'(z)}
& &
v'(z)=\frac{-C z^3 \text{csch}(r(z)) \sqrt{32 C^2 z^6+15 f(z) \cosh (2 r(z))-6 f(z) \cosh (4 r(z))+f(z) \cosh (6 r(z))-10 f(z)} }{8\left(C^2 z^6 f(z)+f(z)^2 \sinh ^6(r(z))\right)} \nonumber\\
& & \hskip 0.5in  \times \left(\sqrt{2 z^2 f(z) r'(z)^2+\cosh (2 r(z))-1}-8 C^2 z^6-8 f(z) \sinh ^6(r(z))\right).
\end{eqnarray}
}
Euler-Lagrange equation of motion for the $r(z)$ from (\ref{AHM-de-Sitter}) obtained as
{\footnotesize
\begin{eqnarray}
\label{EL-EOM-r(z)}
& &
\frac{\sinh ^2(r(z)) }{2 z^4 \left(z^2 r'(z)^2-\sinh ^2(r(z)) v'(z)
   \left(f(z) v'(z)+2\right)\right) \sqrt{r'(z)^2-\frac{\sinh ^2(r(z)) v'(z) \left(f(z) v'(z)+2\right)}{z^2}}}  \Biggl(z r'(z) \sinh ^2(r(z)) \nonumber\\
   & &
   \left(\left(z f'(z)+2 f(z)\right) v'(z)^2+2 v'(z) \left(z f(z) v''(z)+2\right)+2 z v''(z)\right)-\sinh ^2(r(z)) v'(z) \left(f(z) v'(z)+2\right) \nonumber\\
   & &\left(3
   f(z) \sinh (2 r(z)) v'(z)^2+2 z^2 r''(z)+6 \sinh (2 r(z)) v'(z)\right)+4 z^2 r'(z)^2 \sinh (2 r(z)) v'(z) \left(f(z) v'(z)+2\right)-4 z^3 r'(z)^3\Biggr)
   =0. \nonumber\\
\end{eqnarray}
}
Substituting $v'(z)$ from (\ref{v'(z)}) into (\ref{EL-EOM-r(z)}) and using $f(z)=1-z^2$, we set $z_s=1$ for simplification, EOM (\ref{EL-EOM-r(z)}) simplifies to the following form
{\footnotesize
\begin{eqnarray}
\label{EL-EOM-r(z)-simp}
& &\hskip -0.3in
\frac{\sinh ^2(r(z)) }{2 z^4 \left(C^2
   z^6-\left(z^2-1\right) \sinh ^6(r(z))\right) \left(z^2 \left(z^2-1\right) r'(z)^2-\sinh ^2(r(z))\right) \sqrt{\frac{\left(z^2-z^4\right) r'(z)^2 \sinh ^6(r(z))+\sinh ^8(r(z))}{C^2
   z^8+\left(z^2-z^4\right) \sinh ^6(r(z))}}}\nonumber\\
   & & \hskip -0.3in \times \Biggl(-2 z^2 r''(z) \sinh ^2(r(z)) \left(C^2 z^6-\left(z^2-1\right) \sinh ^6(r(z))\right)+r'(z) \left(2 z \sinh ^8(r(z))-4 C^2 z^7 \sinh ^2(r(z))\right)\nonumber\\
   & &\hskip -0.3in +r'(z)^2 \left(C^2 z^8 \sinh
   (2 r(z))-8 z^2 \left(z^2-1\right) \sinh ^7(r(z)) \cosh (r(z))\right)+r'(z)^3 \left(4 z^3 \left(z^2-1\right)^2 \sinh ^6(r(z))-2 C^2 z^9\right)\nonumber\\
   & & +6 \sinh ^9(r(z)) \cosh (r(z))\Biggr)=0.
\end{eqnarray}
}
The above equation is difficult to solve. One trivial solution of (\ref{EL-EOM-r(z)-simp}) is
\begin{eqnarray}
\label{soln-r(z)}
r(z)=0.
\end{eqnarray}
From equation (\ref{AHM-de-Sitter}), we can see that when $r(z)=0$\footnote{Same solution $r(z)=0$ also appeared in \cite{Massless-Gravity} in the computation of area of Hartman-Maldacena surface. See \cite{GB-3} for the similar solution, in our case embedding is $r(z)$ whereas in \cite{GB-3}, embedding is $r(\mu)$, $\mu$ being the angle.} then $A_{\rm HM}^{\rm de-Sitter}=0$\footnote{See \cite{Chapman} for the discussion of complexity of de-Sitter spaces.}, and hence area of Hartman-Maldacena surface is
\begin{eqnarray}
\label{AHM-de-Sitter-on-shell}
& & A_{\rm HM}^{\rm de-Sitter}=0,
\end{eqnarray}
we see that area of the Hartman-Maldacena surface vanishes and hence
\begin{eqnarray}
\label{SHM-dS}
& & S_{\rm HM}^{\rm de-Sitter} = \frac{A_{\rm HM}^{\rm de-Sitter}}{4 G_N^{(5)}} =0.
\end{eqnarray}

{\bf Cosmological Island Surface Entanglement Entropy}:
Area of the island surface parametrized by $t=constant$, $z=z(r)$ 
can be obtained from the induced metric in terms of embedding ($z=z(r)$) and its derivative using (\ref{de-Sitter-metric-SdS}) and the final result is
\begin{eqnarray}
\label{action-dS-IS}
A_{\rm IS}^{\rm de-Sitter}= V_2 \int_{- \rho}^{\rho} dr  \Biggl(\frac{\sinh^{2}(r)}{z(r)^2} \sqrt{1+\frac{\sinh^{2}(r) z'(r)^2}{z(r)^2 (1-z(r)^2)}}\Biggr).
\end{eqnarray}
For the de-Sitter patch, $f(z)=1-\left(\frac{z}{z_s}\right)^2$, we have taken $z_s=1$ in (\ref{action-dS-IS}) for calculation simplification. Therefore $f(z)\geq 0$ if $0<z<1$. Euler-Lagrange equation of motion for the embedding $z(r)$ from (\ref{action-dS-IS}) turns out to be:
{\footnotesize
\begin{eqnarray}
\label{EOM-dS-IS}
& & \hskip -0.3in
\frac{\sinh ^2(r) \sqrt{\frac{-\sinh ^2(r) z'(r)^2+z(r)^4-z(r)^2}{z(r)^2 \left(z(r)^2-1\right)}} }{\left(z(r) \sinh ^2(r)
   z'(r)^2-z(r)^5+z(r)^3\right)^2}\Biggl(z(r) \sinh ^2(r) z'(r)^2+3
   \sinh ^3(r) \cosh (r) z'(r)^3 -z(r)^4 \sinh (r) \left(\sinh (r) z''(r)+4 \cosh (r) z'(r)\right)\nonumber\\
   & &+z(r)^2 \sinh (r) \left(\sinh
   (r) z''(r)+4 \cosh (r) z'(r)\right)+2 z(r)^7-4 z(r)^5+2 z(r)^3\Biggr)
   =0.
\end{eqnarray}
}
In general, it is not easy to solve the above equation. Interestingly, there is a $z(r)=1$ solution to the above differential equation which is nothing but de-Sitter horizon assumed earlier($z_s=1$)\footnote{This can also be verified from the terms inside the open bracket of (\ref{EOM-dS-IS}). Apart from $(2 z(r)^7-4 z(r)^5+2 z(r)^3)$, every term contains the derivative of $z(r)$. For $z(r)=1$, $ 2 z(r)^7-4 z(r)^5+2 z(r)^3=0$ and hence $z(r)=1$ satisfies (\ref{EOM-dS-IS}). There are other two possibilities $z(r)=-1$ and $z(r)=0$ but entanglement entropy (\ref{action-dS-IS}) is negative for $z(r)=-1$ and divergent for $z(r)=0$ and hence these are non-physical.} and it satisfies the Neumann boundary condition on the branes and hence the solution for the cosmological island surface is
\begin{eqnarray}
\label{soln-z(r)-dS-i}
z(r)=1.
\end{eqnarray}
One can arrive at the same conclusion by requiring the well-defined variational principle of (\ref{action-dS-IS}) and imposing Neumann boundary condition on the branes similar to the discussion in section \ref{PC-TEBH} which requires
\begin{eqnarray}
\label{NBC-dS}
\frac{\sinh ^4(r) z'(r)}{z(r)^4 \left(1-z(r)^2\right) \sqrt{\frac{\sinh ^2(r) z'(r)^2}{z(r)^2 \left(1-z(r)^2\right)}+1}}=0.
\end{eqnarray}
When we impose $z'(r=\pm \rho)=0$ then the minimal surface is the horizon, i.e., $z(r)=1$ \cite{GB-3}. On substituting $z(r)=1$ in (\ref{action-dS-IS}), we obtain the minimal area of the cosmological island surface for the de-Sitter patch as given below
\begin{eqnarray}
\label{AIS-simp-dS}
A_{\rm IS}^{\rm de-Sitter}= V_2 \int_{-\rho}^{\rho} dr \sinh^{2}(r)=V_2 \left(\sinh (\rho ) \cosh (\rho )-\rho\right).
\end{eqnarray}
Entanglement entropy contribution of cosmological island surface is
\begin{eqnarray}
\label{SIS-dS}
S_{\rm IS}^{\rm dS}=\frac{2 A_{\rm IS}^{\rm de-Sitter}}{4 G_N^{(5)}}= 2 S_{\rm dS}^{\rm thermal}.
\end{eqnarray}
Additional numerical factor ``2'' is coming due to second cosmological island surface on thermofield double partner side (shown in Fig. \ref{SdS-CP}). We get the Page curve of de-Sitter patch by plotting (\ref{SHM-dS}) and (\ref{SIS-dS}) from wedge holography. We will get a flat Page curve in this case similar to \cite{GB-3}.

Let us summarize the results of this section. It was argued in \cite{GB-3,Massless-Gravity} that in wedge holography without DGP term, the black hole horizon is the only extremal surface and the Hartman-Maldacena surface does not exist and hence one expects the flat page curve. We also see that when we compute the entanglement entropies of island surfaces of AdS, Schwarzschild, and de-Sitter black holes then minimal surfaces turn out to be horizons of the AdS or Schwarzschild or de-Sitterblack holes. As a curiosity, we computed entanglement entropies of Hartman-Maldacena surfaces for the parametrization $r(z)$ and $v(z)$ used in the literature and we found non-trivial linear time dependence for the AdS and Schwarzschild black holes whereas Hartman-Maldacena surface entanglement entropy turns out to be zero for the de-Sitter black hole. Therefore we obtain the flat Page curve for the de-Sitter black hole not for the AdS and Schwarzschild black holes due to the non-zero entanglement entropy of Hartman-Maldacena surfaces. 
The theme of the paper is not to discuss whether we get a flat Page curve or not. The paper aimed to construct a ``multiverse'' in Karch-Randall braneworld which we did in section \ref{Multiverse-section} and check the formula given in (\ref{Island-Formula-n-BHs}). We saw in subsection \ref{PC-TEBH} that (\ref{Island-Formula-n-BHs}) is giving consistent results.

{\bf Comment on the Wedge Holographic Realization of Schwarzschild de-Sitter Black Hole with Two Karch-Randall Branes}:
In subsection \ref{IP-SdS}, we performed our computation of the Schwarzschild and de-Sitter patches separately. There is one more way by which we may get the Page curve of the Schwarzschild de-Sitter black hole. We summarize the idea below:
\begin{itemize}
\item Consider two Karch-Randall branes $Q_1$ and $Q_2$ such that one of which contains Schwarzschild de-Sitter black hole and the other one act as a bath to collect the radiation\footnote{In this case, Hawking radiation will not be a suitable term because when Schwarzschild de-Sitter black hole as whole emits radiation then observer may not distinguish between Hawking radiation emitted by Schwarzschild patch and Gibbons-Hawking radiation emitted by de-Sitter patch \cite{GH-radiation}.}.

\item Suppose the bulk metric has the following form:
\begin{equation}
\label{metric-SdS-induced}
ds^2 = g_{\mu \nu} dx^\mu dx^\nu=dr^2+g(r) h_{ij}^{\rm SdS} dy^i dy^j=dr^2+g(r) \Biggl(\frac{\frac{dz^2}{f(z)}-f(z) dt^2+\sum_{i=1}^{2}dy_i^2}{z^2} \Biggr),
\end{equation}
where $f(z)=1-\frac{2 M}{z}-\frac{\Lambda}{3}z^2$ in $d=4$.
\item Next step is find out $g(r)$ by solving Einstein equation (\ref{Einstein-equation}).

\item After getting the solution, one needs to ensure that bulk metric (\ref{metric-SdS-induced}) must satisfy the Neumann boundary condition (\ref{NBC}) at $r=\pm \rho$.

\item One also needs to check what kind of theory exists at the defect i.e., it is CFT or non-CFT, to apply the Ryu-Takayanagi formula.

\item If we are successfully checked the above points then we can obtain the Page curve of the Schwarzschild de-Sitter black hole by computing the areas of Hartman-Maldacena and island surfaces\footnote{In this setup, the notion of ``island'' may become problematic because we will be talking about the island in the interior of Schwarzschild de-Sitter black hole. Since SdS black hole has two horizons, therefore it may cause trouble to say whether the ``island'' is located inside the black hole horizon or the de-Sitter horizon. Therefore it will be nice to follow the setup with two black holes and two baths. See \cite{Gopal+Nitin} for non-holographic approach.}. 
\end{itemize}
The above discussion is just a ``mathematical idea''. Since we have three possible branes: Minkowski, de-Sitter and anti de-Sitter \cite{mismatched-branes}. There is no brane with the induced metric defined in the open bracket of (\ref{metric-SdS-induced}). Further, we have AdS/CFT correspondence or dS/CFT correspondence, or flat space holography. There is no such duality that states the duality between CFT and bulk which has the form of a Schwarzschild de-Sitter-like structure. There will be no defect description due to the aforementioned reason and hence no ``intermediate description'' of wedge holography. Therefore we conclude that one can model Schwarzschild de-Sitter black hole from wedge holography with two copies of wedge holography in such a way that one part defines Schwarzschild patch and the other part defines de-Sitter patch\footnote{See \cite{Gopal+Nitin,Baek} for non-holographic model.}.

\section{Application to Grandfather Paradox}
\label{AGP}
This section states the ``grandfather paradox'' and its resolution in our setup.

``Grandfather paradox'' says that Bob can not travel back in time. Because if he can travel back in time, he can land in another universe where he can kill his grandfather. If Bob's grandfather is dead in another universe, then he will not exist in the present \cite{GP}.

Now let us see how this problem can be avoided in our setup. We discussed in sections \ref{AdS-multiverse} and \ref{de-Sitter-multiverse} that a multiverse consists of $2n$ Karch-Randall branes, which we call ``universes''. The geometry of these branes is AdS and de-Sitter spacetime in sections \ref{AdS-multiverse} and \ref{de-Sitter-multiverse}. In all the setups, all ``universes'' are connected at the ``defect'' via transparent boundary condition. Transparent boundary condition guarantees that all these universes are communicating with each other.
\begin{figure}
\begin{center}
\includegraphics[width=0.8\textwidth]{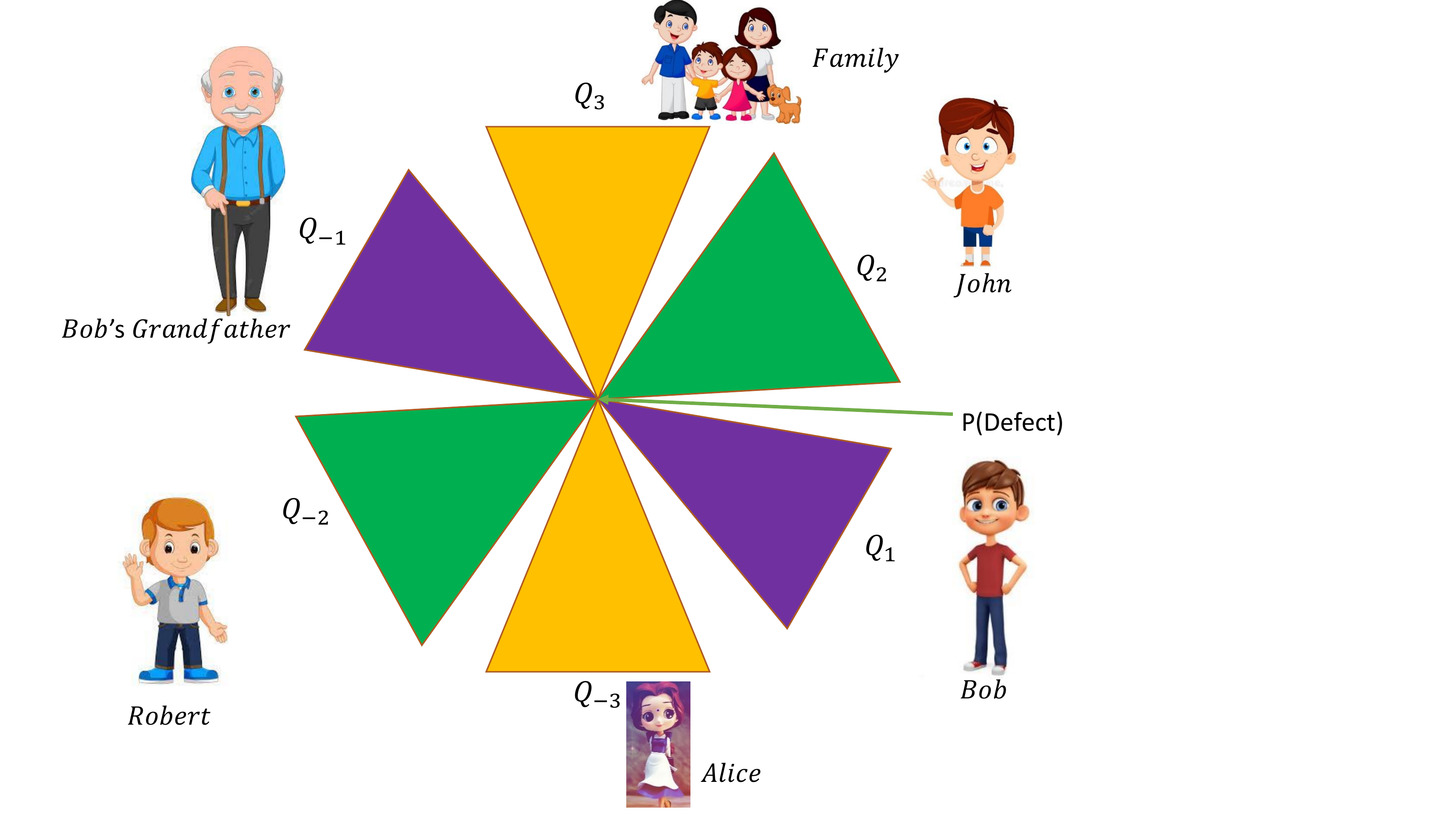}
\end{center}
\caption{Different universes $Q_{-1,-2,-3,1,2,3}$ where different people are living.}
\label{RGP}
\end{figure}

Suppose Bob lives on $Q_1$ and his grandfather lives on $Q_2$. Then to avoid the paradox, Bob can not travel to $Q_2$, but he can travel to $Q_{-2}$, $Q_{-3}$ etc. where he can meet Robert and Alice. Hence ``grandfather paradox'' can be resolved in this setup. Further traversable wormhole solution is also possible \cite{Maldacena-TWH}. This discussion is consistent with ``many world theory'' where ``grandfather paradox'' has been resolved using the similar idea.

\section{Conclusion}
\label{Conclusion}
In this work, we propose the existence of a multiverse in the Karch-Randall braneworld using the idea of wedge holography. Multiverse is described in the sense that if we talk about $2n$ universes, then those will be represented by Karch-Randall branes embedded in the bulk. These branes will contain black holes or not that can be controlled by gravitational action. We studied three cases:
\begin{itemize}
\item We constructed mutiverse from $d$-dimensional Karch-Randall branes embedded in $AdS_{d+1}$ in section \ref{AdS-multiverse}. The geometry of these branes is $AdS_d$. In this case, the multiverse consists of $2n$ anti de-Sitter branes and all are connected to each other at the defect via transparent boundary conditions. Multiverse consists of AdS branes exists forever once created.

\item We constructed multiverse from $d$-dimensional de-Sitter spaces on Karch-Randall branes embedded in $(d+1)$-dimensional bulk $AdS_{d+1}$ in \ref{de-Sitter-multiverse}. Multiverse made up of $2n$ de-Sitter branes has a short lifetime. All the de-Sitter branes in this setup should be created and annihilated at the same time. Defect CFT is a non-unitary conformal field theory because of dS/CFT correspondence.

\item We also discussed why it is not possible to describe multiverse as a mixture of $d$-dimensional de-Sitter and anti de-Sitter spacetimes in the same bulk in section \ref{AdS+dS-Multiverse}. We can have the multiverse with anti de-Sitter branes ($M_1$) or de-Sitter branes ($M_2$) but not the mixture of the two. Because AdS branes intersect at ``time-like'' boundary and de-Sitter branes intersect at ``space-like'' boundary of the bulk $AdS_{d+1}$. Universes in $M_1$ can communicate with each other, similarly, $M_2$ consists of communicating de-Sitter branes but $M_1$ can't communicate with $M_2$.  
\end{itemize}

We look for the possibility of whether we can resolve the information paradox of multiple black holes simultaneously or not. This can be done by constructing a multiverse in such a way the $n$ Karch-Randall branes will contain black holes, and Hawking radiation of these black holes will be collected by a $n$ gravitating baths. In this case, we obtain linear time dependence from the Hartman-Maldacena surfaces, and the constant value will be $2 S^{{i=1,2,..,n}, \ \rm thermal}_{\rm BH}$ which is coming from $n$ island surfaces. 

As a consistency check of the proposal, we calculated the Page curves of two black holes for $n=2$ multiverse. We assumed that black hole and bath systems between $- 2 \rho \leq r \leq 2 \rho$ and $-  \rho \leq r \leq \rho$. In this case, we found that entanglement entropy contribution from the Hartman-Maldacena surfaces has a linear dependence on time for the AdS and Schwarzschild black holes and it is zero for the de-Sitter black hole, whereas island surfaces contributions are constant. Therefore this reproduces the Page curve. Using this idea, we obtain the Page curve of Schwarzschild de-Sitter black hole and one can also do the same for Reissner-Nordstr\"om de-Sitter black hole. This proposal is helpful in the computation of the Page curve of black holes with multiple horizons from wedge holography. We also discussed the possibility of getting a Page curve of these black holes using two Karch-Randall branes, one as a black hole and the other as a bath. In this case, there will be an issue in defining the island surface and identifying what kind of radiation we are getting. For example, when a Karch-Randall brane consists of black hole and cosmological event horizons, i.e., Schwarzschild de-Sitter black hole on the brane, the observer collecting the radiation will not be able to identify clearly whether it is Hawking radiation or Gibbons-Hawking radiation.

We checked our proposal for very simple examples without DGP term on the Karch-Randall branes, but one can also talk about massless gravity by adding the DGP term on the Karch-Randall branes \cite{Massless-Gravity}. In this case, tensions of the branes will recieve correction from the extra term in (\ref{NBC-DGP}).
 Further, we argued that one could resolve the ``grandfather paradox'' using this setup where all universes communicate via transparent boundary conditions at the interface point. To avoid the paradox, one can travel to another universe where his grandfather is not living, so he can't kill his grandfather. We have given a qualitative idea to resolve the ``grandfather paradox'' but detailed analysis requires more research in this direction using wedge holography.

\section*{Acknowledgements}
The author is supported by a Senior Research Fellowship (SRF) from the Council of Scientific and Industrial Research (CSIR), Govt. of India. It is my pleasure to thank Aalok Misra, who motivated me to work on the entanglement stuff, and for his blessings. We would also like to thank Juan Maldacena, Andreas Karch, Kostas Skenderis and Tadashi Takayanagi for very helpful discussions and comments. This research was also supported in part by the International Centre for Theoretical Sciences (ICTS) for the program ``Nonperturbative and Numerical Approaches to Quantum Gravity, String Theory and Holography'' (code:ICTS/numstrings-2022/8). Various conferences/workshops; e.g., {{\it Mysteries of Universe-I (Institute Lecture Series)} and {\it Indian Strings Meeting 2021}} at Indian Institute of Technology Roorkee, Roorkee, India; {\it Applications of Quantum Information in QFT and Cosmology} at the University of Lethbridge, Canada; {\it Kavli Asian Winter School (KAWS) on Strings, Particles and Cosmology (Online)} at International Centre for Theoretical Sciences (ICTS) Bangalore, India (code:ICTS/kaws2022/1); {\it Reconstructing the Gravitational Hologram with Quantum Information} at Galileo Galilei Institute for Theoretical Physics, Florence, Italy; {\it Quantum Information in QFT and AdS/CFT-III} at Indian Institute of Technology Hyderabad, India; helped me to learn about the information paradox and related stuff. I am very thankful to the speakers and organizers of these conferences because I learned about the subject from these conferences.

\end{document}